\documentclass[]{article}
\usepackage{lineno,hyperref}
\usepackage{amsmath,amsfonts}
\usepackage[abs]{overpic}
\usepackage{bm}
\usepackage{pstricks,soul}
\usepackage[affil-it]{authblk}
\newcommand{\R}{\mathbb{R}}
\renewcommand{\O}{\mathcal{O}}
\newcommand{\E}{\mathcal{E}}

\newcommand{\Dt}{\Delta {t}}
\newcommand{\Ad}{A}
\newcommand{\Wz}{z}

\newcommand{\Crand}{C_z}
\newcommand{\Cchar}{C_{s}}
\newcommand{\Cr}{C_{r}}
\newcommand{\Ct}{C_{\tau}}
\newcommand{\Ca}{C_{a}}
\newcommand{\barS}{s}
\newcommand{\xf}{x}

\newcommand{\vf}{v}

\newcommand{\target}{x^\tau}

\begin{document}

\title{Robust Design Optimization for Egressing Pedestrians in Unknown Environments}

\author[1]{Emiliano Cristiani}
\author[1,*]{Daniele Peri}
\affil[1]{Istituto per le Applicazioni del Calcolo, Consiglio Nazionale delle Ricerche, \\ Via dei Taurini 19, 00185 Rome, Italy}
\affil[*]{Corresponding author, d.peri@iac.cnr.it}
\date{}

\maketitle

\begin{abstract}
	In this paper, we deal with a size-variable group of pedestrians moving in a unknown confined environment and searching for an exit. Pedestrian dynamics are simulated by means of a recently introduced microscopic (agent-based) model, characterized by an exploration phase and an egress phase. First, we study the model to reveal the role of its main parameters and its qualitative properties. Second, we tackle a robust optimization problem by means of the Particle Swarm Optimization method, aiming at reducing the time-to-target by adding in the walking area multiple obstacles optimally placed and shaped. Robustness is sought against the number of people in the group, which is an uncertain quantity described by a random variable with given probability density distribution.
\end{abstract}


\section{Introduction}\label{sec:introduction}
In this paper, we are concerned with modeling and control of pedestrian dynamics. 
More precisely, we consider the case of a group formed by an uncertain number of people who enter a unknown confined environment and try to reach a target (e.g.\ a door). We assume that agents are not informed about the position of the target, so they need to explore the environment first. Moreover, the target is recognized as such only when people are close enough to it (cf.\ papers \cite{carrillo1501.07054, cirillo2013PhysA, guo2012TRB} where zero-visibility conditions are considered). 
As a guide scenario, we can think at a flow of people entering a ticket office of a large museum and then looking for the beginning of the exhibition path. Metro stations, airports and malls can also be reasonable test cases.
In that case, we claim that people exhibit two opposite tendencies: on the one hand, they tend to spread out in order to explore the environment efficiently; on the other hand, they follow the group mates hopeing that others have already found a way to the target; see, e.g., experimental papers \cite{albi2016SIAP, haghani2017AB, dyer2009PTRSB} where this kind of social influence is investigated. 

We aim at modifying the design of the environment by \emph{adding} obstacles in order to facilitate the research of the target. This can be interpreted as a reverse application of Braess's paradox, originally formulated in the context of traffic flow on road networks \cite{braess2005TS, hughes2003ARFM}, which states that increasing the network connection (e.g., creating a new road) can actually decrease the overall flow of the network.

Several papers investigate the effectiveness of the Braess's paradox \textit{in silico}, reporting the effect of additional obstacles placed in the walking area. A manual placing is performed, e.g., in  \cite{escobar2003LNCS, frank2011PA, helbing2005TS, hughes2002TRB, matsuoka2015, twarogowska2014AMM}, while an optimization algorithm is used, e.g., in \cite{cristiani2015SIAP, cristiani2017AMM, jiang2014PLOS, johansson2007, shukla2009, zhao2017PhysA}.
Such a bottom-down crowd control strategy pursued by the smart deployment of obstacles can be very effective in the case of very large crowds or emergency situations, and, more in general, whenever communications from supervisors to the crowd are difficult.

Given that, one of the most serious criticisms to this approach is that \emph{the optimal shape and position of the obstacles strongly depend on the initial position and the number of people}, meaning that a given design could be advantageous in some situations and disadvantageous in another ones. This issue motivates the \textit{robust} optimal control problem studied in this paper. We assume that the number of people is an uncertain quantity described by a random variable with given probability density distribution and we compute the optimal obstacles design in accordance with this distribution. As a consequence, we expect that the suggested design allows decreasing the time-to-target in the most likely scenarios, simultaneously limiting the disadvantages in the less probable cases.

The observation of the final results can provide/confirm some insights and guidelines for the distribution of obstacles into an environment.

\paragraph{Methods}
A very good source of references about evacuation models can be found in \cite{abdelghany2014EJOR}, where methods both with and without optimal planning search are discussed.

In this paper, pedestrian dynamics are simulated by the model introduced in \cite{albi2016SIAP} which adopts a microscopic (agent-based) point of view, i.e.\ it tracks every single pedestrian individually by means of a system of stochastic ordinary differential equations, in the same spirit of the classical social force model \cite{helbing1995PRE}. 
The model is specifically conceived to deal with unknown environments, possibly with limited visibility, and it is characterized by an \textit{exploration phase} and an \textit{egress phase}. 
This is achieved by a suitable combination of repulsion, alignment, and self-propulsion social forces. 
Moreover, people interact with each other if they are closer than a certain distance, thus avoiding unnatural all-to-all interactions.

A rigorous application of optimization algorithms in the context of pedestrian dynamics is rarely adopted. Commonly, a simple systematic or comparative analysis between two or more alternative solution is proposed, but only in a few studies a mathematical programming problem is formulated. 
Examples are \cite{cristiani2017AMM} and \cite{giacomini2017ACE}, both presenting the solution of a deterministic optimization problem, where all the involved parameters are fixed statically but the optimizing variables. Unfortunately, in real-life situations some of the environmental conditions are substantially unknown, and some hypotheses can be only argued through probabilistic assumptions, defining what is commonly called {\em Robust Design Optimization (RDO) problem}. To the authors knowledge there are no examples of RDO applications in the field, while in the aeronautic, automotive and electronics fields, to name a few, RDO can be considered as {\em common practice}.
A good review paper is \cite{Chatterjee2017}, where also the concept of the approximation techniques applied in the optimization context is illustrated.
For a paper with an application of RDO with real-life data, we can refer, among the others, to \cite{peri2016OcEng}.

\medskip

\paragraph{Paper organization}
In Section \ref{sec:model}, we detail the model and the way in which the interactions with obstacles are treated.
In Section \ref{sec:modelstudy}, we study the model, with special emphasis to the role of its main parameters and its statistical properties. We also discuss the choice of the objective function which will be used in the optimization problem.
In Section \ref{sec:shapeoptimization}, we discuss how the obstacles are parameterized, in order to introduce suitable control variables for the optimization problem. The optimization problem is then solved on two simple scenarios, i.e.\ a room with single entrance and multiple entrances, respectively. 
Finally, in Section \ref{sec:conclusions} we sketch some conclusions. 

\section{Model and obstacle's management}\label{sec:model}
Let us assume that $N$ pedestrians are free to move in a bounded walking area $\Omega\subset\R^2$ except for the areas occupied by obstacles. 
Assume also, for simplicity, that there are only one entrance $\E$ and one target $\target$ (exit) both on the boundary of $\Omega$, and one obstacle $\O$ inside $\Omega$ (extension to multiple entrances, exits, and obstacles is straightforward).
To define the exit's visibility area, we consider the set $\Sigma$, with $\target\in\Sigma\subset\Omega$, and we assume that the target is completely visible from any point belonging to $\Sigma$ and completely invisible (or simply not recognized as such) from any point belonging to $\Omega\backslash\Sigma$. 

For every $i=1,\ldots,N$, let $(\xf_i(t),\vf_i(t))\in\R^2\times\R^2$ denote position and velocity of the agents at time $t\geq 0$. 

\medskip

The dynamics are described by the following system of stochastic ordinary differential equations:
\begin{equation}\label{eq:model}
\left\{
\begin{array}{l}
\dot{x}_i = \vf_i,\\ [1.5mm] 
\dot{v}_i = \Ad(\xf_i,\vf_i) + \sum_{j=1}^{N} H(\xf_i,\vf_i,\xf_j,\vf_j)
\end{array}
\right. ,
\qquad i = 1, \dots, N,
\end{equation}
where

\begin{itemize}

\item $\Ad$ is a self-propulsion term, given by the relaxation toward a random direction (exploration phase) or the relaxation toward a unit vector pointing to the target (egress phase), plus a term which translates the tendency to reach a given characteristic speed $\barS \geq 0$ (modulus of the velocity). In formulas,
\begin{multline} \label{eq:A}
\Ad(x,v) :=  \chi_{\textup{targ}}(x)\Ct\left(\frac{\target - x}{\|\target - x\|} - v\right)+
(1-\chi_{\textup{targ}}(x)) \Crand(\Wz-v) +\\
 \Cchar(\barS^2-\|v\|^2)v
\end{multline}
where 
$$
\chi_{\textup{targ}}(x):=\left\{
\begin{array}{ll}
1, & x\in\Sigma, \\
0, & \text{otherwise,}
\end{array}
\right.
$$ 
$\Wz$ is a two-dimensional random vector with normal distribution $\mathcal N(0,1)$, and $\Crand$, $\Ct$, $\Cchar$ are positive constants.

\item
$H$ is an interaction term which embeds an isotropic (all around) metri\-cal short-range repulsion force directed against close group mates (to avoid collisions) and, if the exit is not visible, additionally accounts for an isotropic metric alignment force. In formulas,
\begin{multline}\label{eq:H}
H(x,v,x',v') := 
-\Cr e^{-\|x'-x\|}\frac{x'-x}{\|x'-x\|} \chi_\textup{inter}(x,x';r_{\textup{rep}}) + \\ 
(1-\chi_{\textup{targ}}(x,x'))\Ca\left(v'-v\right) \chi_\textup{inter}(x,x';r_{\textup{align}})
\end{multline}
where 
$$
\chi_{\textup{inter}}(x,x';r):=\left\{
\begin{array}{ll}
1, & \|x-x'\|<r, \\
0, & \text{otherwise,}
\end{array}
\right. 
$$ 
for given positive constants $r_{\textup{rep}}$, $r_{\textup{align}}$, $\Cr, \Ca$.
Note that, once the summations over the agents $\sum_j H$ are done, the alignment term models the tendency of the followers to relax toward the average velocity of the closer agents.
Note also that no attraction force is considered here, since the alignment itself is fully able to describe the social influence and trigger the ``herding effect'', cf.\ \cite{couzin2002JTB}.

\end{itemize}

The system \eqref{eq:model} is numerically approximated by means of the explicit Euler scheme with time step $\Dt=0.1$.

\paragraph{Boundaries}
Pedestrians are confined in the domain $\Omega\backslash\O$, therefore a special treatment of the boundaries is needed. For the external boundary $\partial\Omega$ we proceed as follows: once an agent is observed to cross the boundary, it is brought back to the closest admissible point and the component of the velocity field which was responsible of the escape is nullified.

Regarding the internal boundaries $\partial\O$, instead, once an agent is found inside an obstacle, it is brought back to the position occupied at the previous time step and both components of the velocity field are nullified.

Finally, we recall that obstacles can be either \textit{transparent} or \textit{opaque}. In the former case, people see and interact with group mates behind obstacles, in the latter they do not. In this paper, we consider the more realistic case of opaque obstacles, employing the numerical technique introduced in \cite[Sect.\ 3.3]{cristiani2017AMM} (see also \cite[Appendix A]{colombo2012M3AS}).

\paragraph{Main limitations}
The model described above suffers from some important limitations, which must be borne in mind when results are discussed.
\begin{enumerate}
\item Agents are assumed not to communicate with each other and not to share information directly.
\item The model does not include some natural tendencies of humans, e.g. the tendency to stay to the right while walking. 
\item Pre-evacuation times are not taken into account. This could have an impact on results if the proposed methodology is employed to find optimal obstacle placement in emergency evacuation scenarios.
\item We assume that people explore the areas with equal probability, regardless of the geometry configuration.
\item No topological interactions are considered. This means that each person interacts with group mates within a certain distance from him/her, and not with a predefined amount of people regardless of their distance (cf.\ \cite{albi2016SIAP}).
\end{enumerate}

It is important to note that most of the limitations listed above are introduced to keep the computational effort low. Indeed, a larger per-run CPU time would make the optimization procedure infeasible.

\section{Model's properties and choice of the objective function}\label{sec:modelstudy}
In this section, we study the model described in Section \ref{sec:model}, paving the way to the optimization problem.

\subsection{Role of $\Crand$ and $\Ca$}
The coupling between random walk and alignment in model \eqref{eq:model} gives rise to interesting phenomena which were little explored so far, cf.\ \cite{albi2016SIAP} and the seminal paper \cite{vicsek1995seminal}. 
The main point is that we observe a continuous fight between the tendency of people to follow their own preferred direction of exploration and the tendency to form a group (social influence). 
To better understand the combination of the two ingredients, we consider the case of a group of people entering a square domain $\Omega=[0,20]^2$ with no exits and no obstacles inside ($\target=\O=\emptyset$). 
At final time $T$, we measure the percentage of domain explored by all people (more precisely, the domain is divided in 10,000 squares and the explored area is computed by counting how many squares were visited at least once by at least one person). 
This study is performed by varying the parameters $\Crand$ and $\Ca$, while keeping fixed the others, see Table \ref{tab:parameters_exploration}. Entrance is located at $\E=(0,5)$ and people enter the domain one at the time, every five time steps. 

\begin{table}[!h]
\caption{Model parameters.}
\label{tab:parameters_exploration}
\begin{center}
\begin{tabular}{|c|c|c|c|c|c|c|c|c|c|}
\hline
$T$ & $N$ & $\Crand$ & $\Ca$ & $\Cchar$ & $C_\tau$ & $C_r$ & $\barS^2$ & $r_{\textup{rep}}$ & $r_{\textup{align}}$ \\
\hline\hline
300 & 50 & 0.05--2.36 & 0--5.9 & 1 & 1 & 2  & 0.5  & 0.4 & 1.2 \\
\hline
\end{tabular}
\end{center}
\end{table}

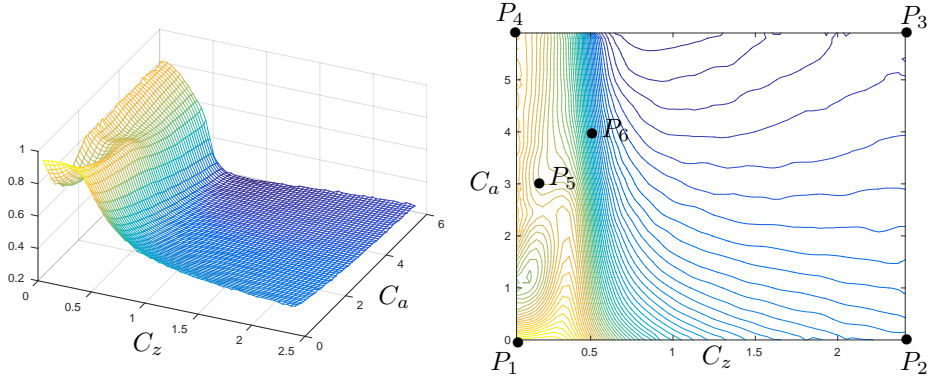
\begin{figure}
\centering
\begin{overpic}[scale=0.45]{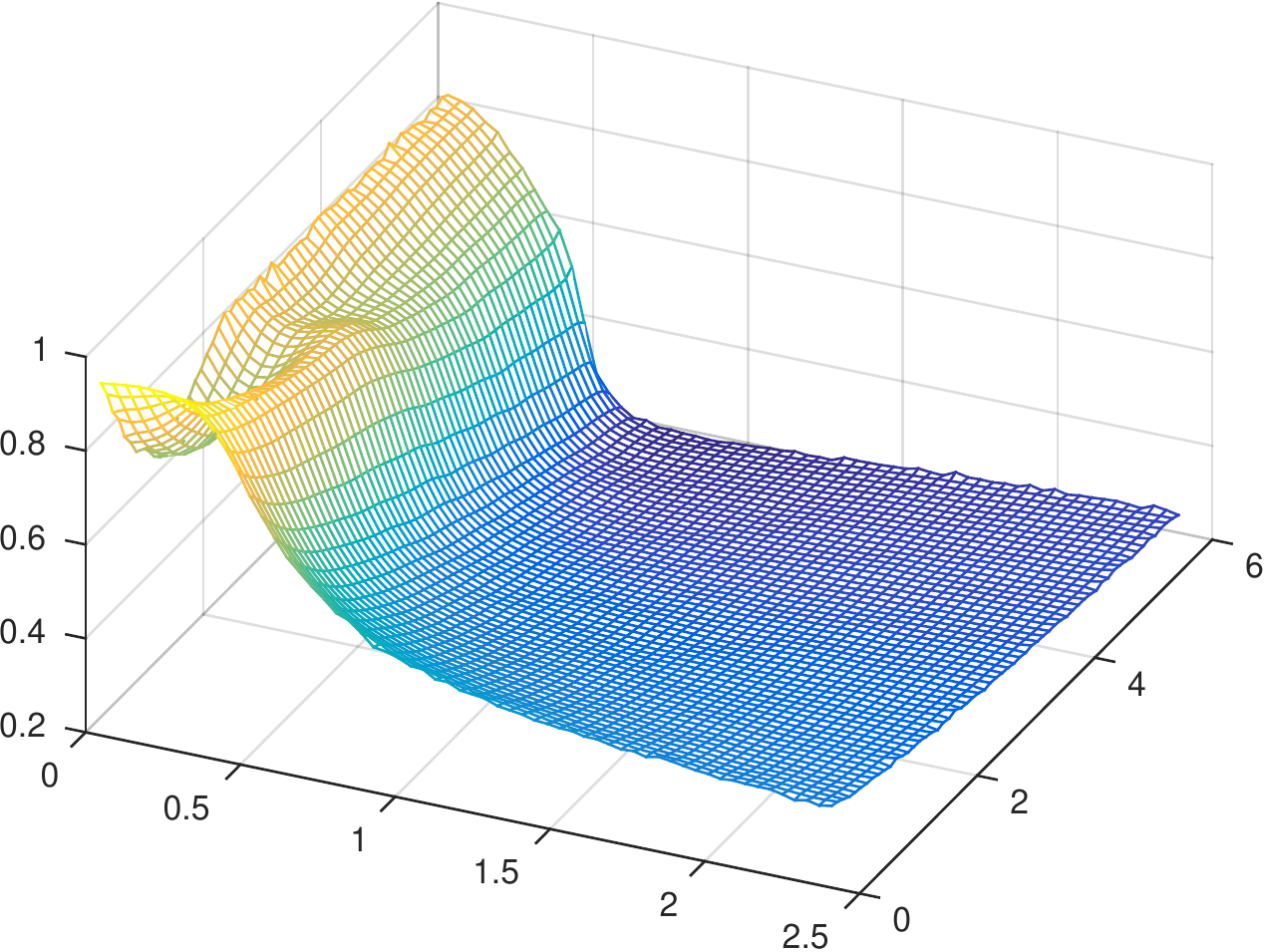}
 \put(140,20){$\Ca$}
 \put(47,2){$\Crand$}
 \end{overpic}
 \qquad
\begin{overpic}[scale=0.45]{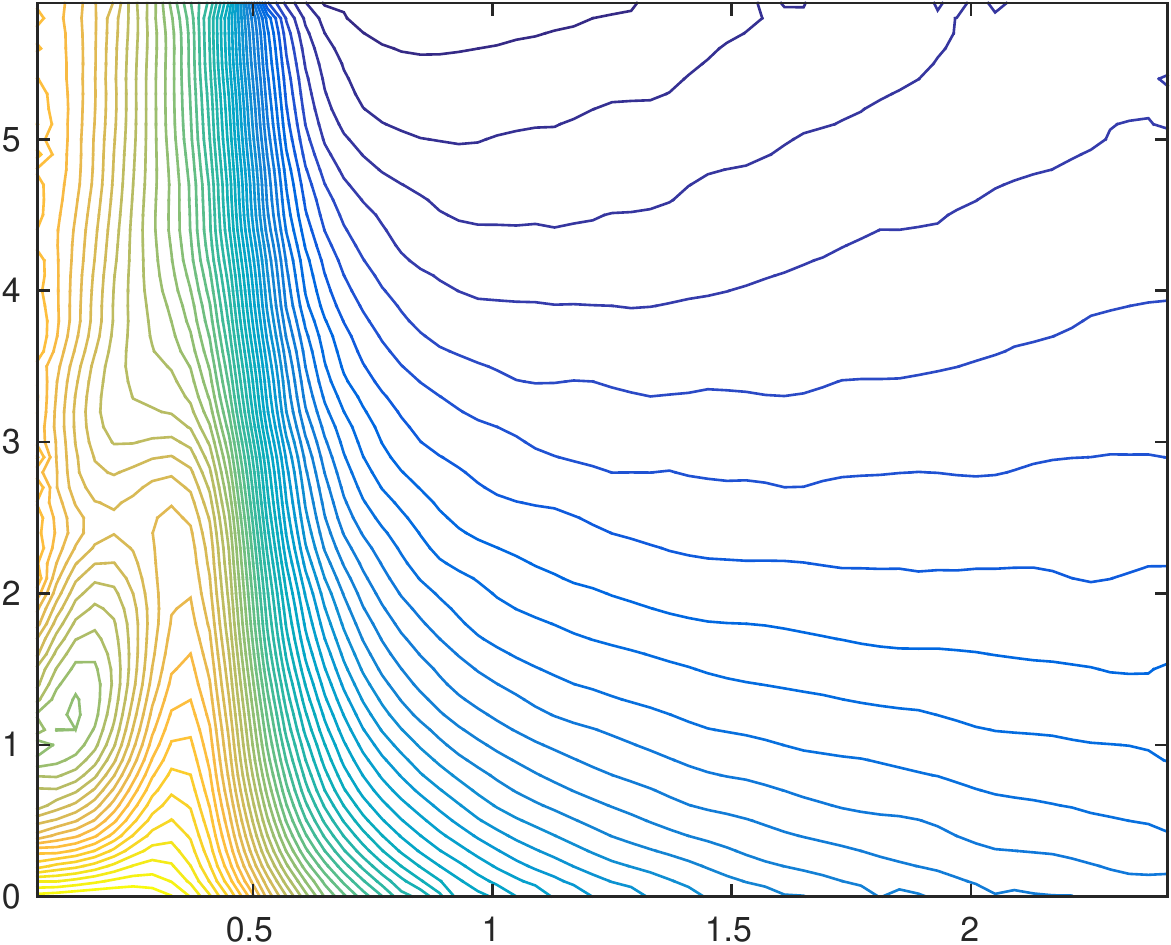}
\put(-5,-5){$P_1$}        \put(3,3){$\bullet$}
\put(150,-5){$P_2$}    \put(150,4){$\bullet$}
\put(150,126){$P_3$}  \put(150,120){$\bullet$}
\put(-3,127){$P_4$}    \put(2,120){$\bullet$}
\put(17,65){$P_5$}    \put(11,63){$\bullet$}
\put(37,83){$P_6$}    \put(31,82){$\bullet$}
\put(75,-3){$\Crand$}   \put(-13,63){$\Ca$}
\end{overpic}
\caption{Average percentage of the explored domain (over 100 runs) as a function of $\Crand$ and $\Ca$. 
$P_1=(0.05,0)$, 
$P_2=(2.36,0)$, 
$P_3=(2.36,5.9)$, 
$P_4=(0.05,5.9)$, 
$P_5=(0.2,3)$,  
$P_6=(0.5,4)$.}
\label{fig:exploration}
\end{figure}

In Fig.\ \ref{fig:exploration}, we show the average percentage of the explored domain (over 100 runs) as a function of $\Crand$ and $\Ca$.
We see that small stochasticity leads to faster exploration; the maximal explored area is reached in the corner $(\Crand,\Ca)=(0.05,0)$, while the minimal one is reached in the opposite corner $(\Crand,\Ca)=(2.4,5.9)$.

\begin{figure}
\centering
\begin{overpic}[width=0.3\textwidth]{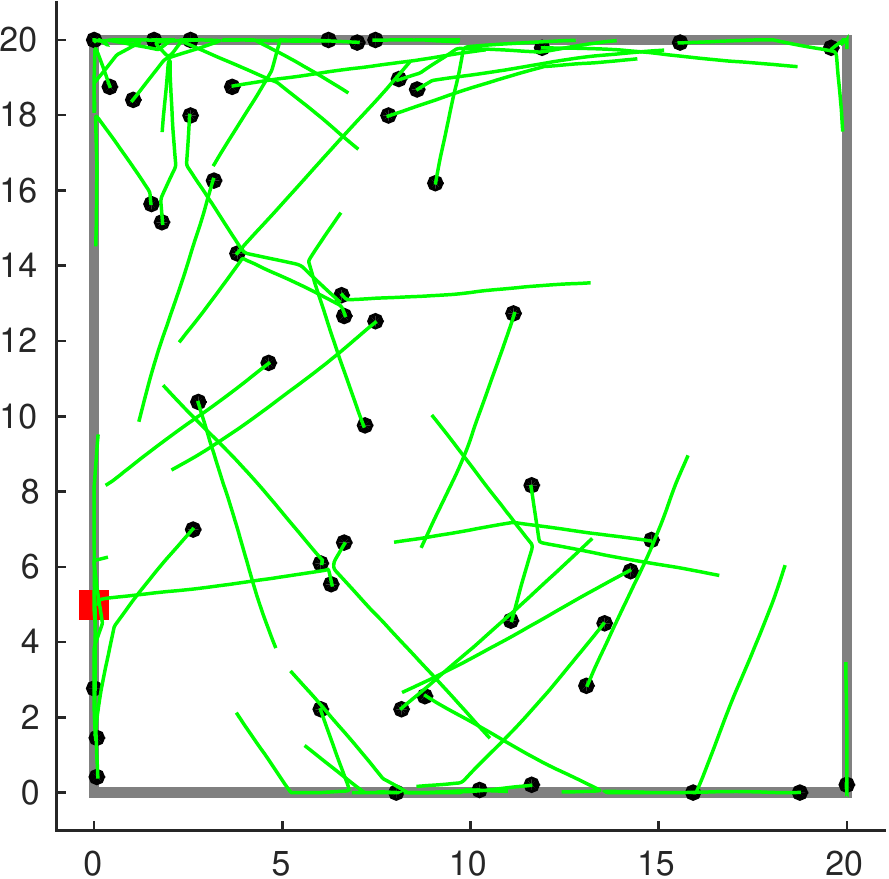}\put(80,80){$P_1$}\end{overpic}
\begin{overpic}[width=0.3\textwidth]{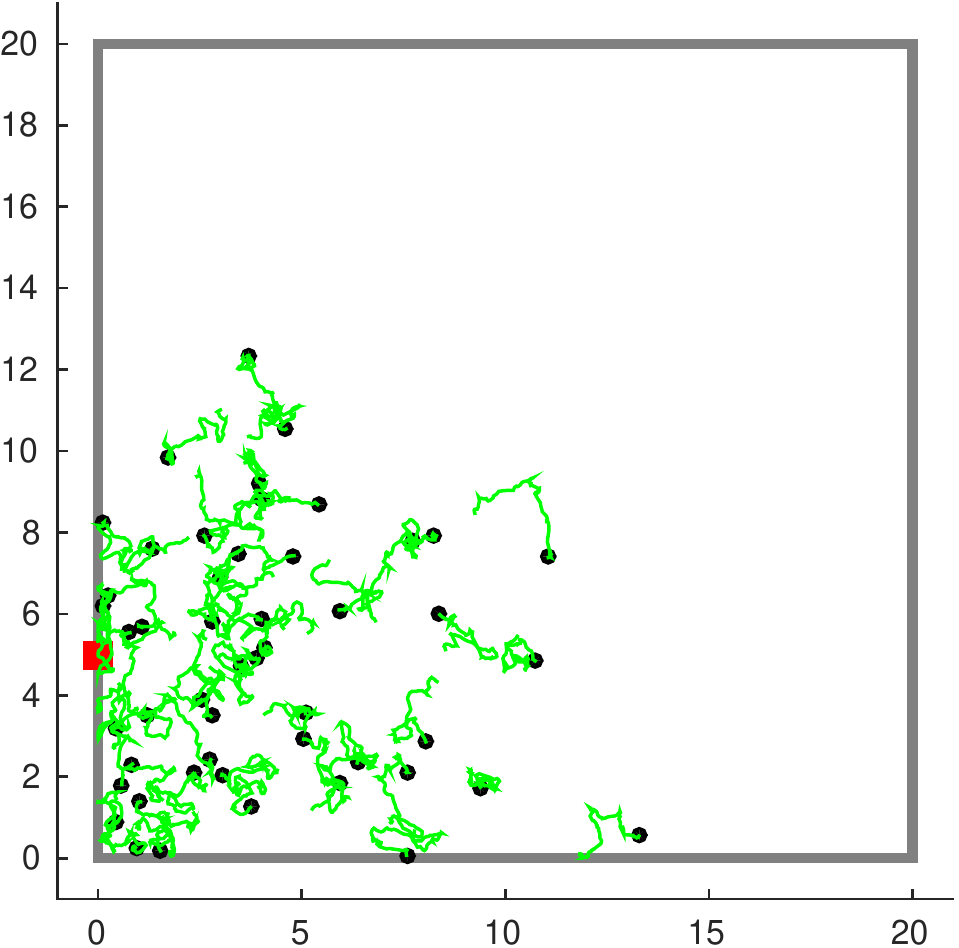}\put(80,80){$P_2$}\end{overpic}
\begin{overpic}[width=0.3\textwidth]{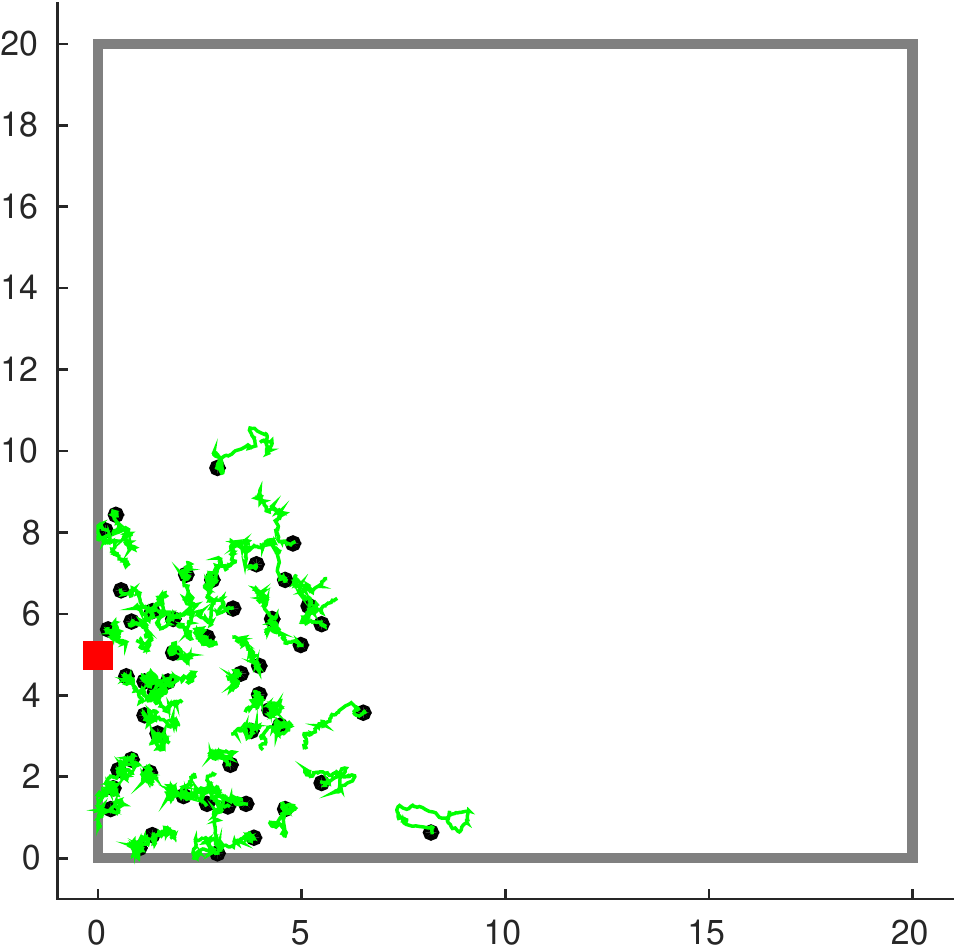}\put(80,80){$P_3$}\end{overpic}
\begin{overpic}[width=0.3\textwidth]{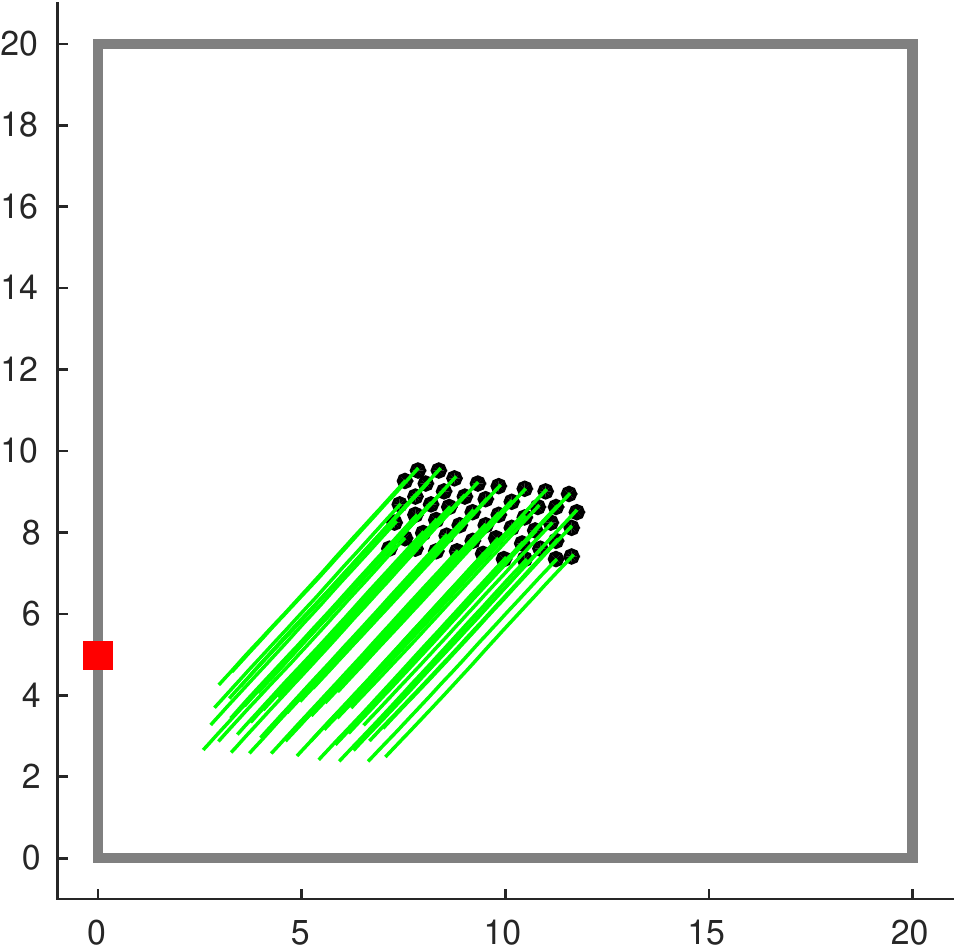}\put(80,80){$P_4$}\end{overpic}
\begin{overpic}[width=0.3\textwidth]{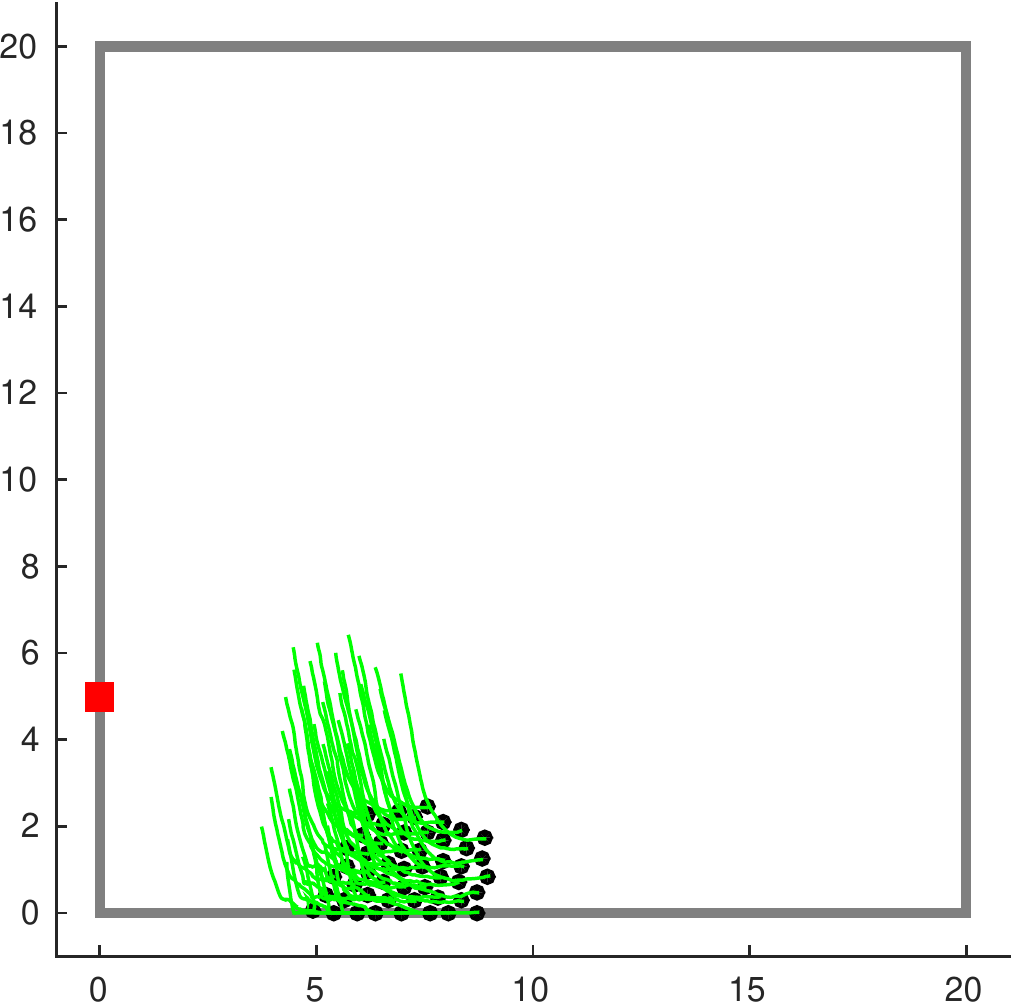}\put(80,80){$P_5$}\end{overpic}
\begin{overpic}[width=0.3\textwidth]{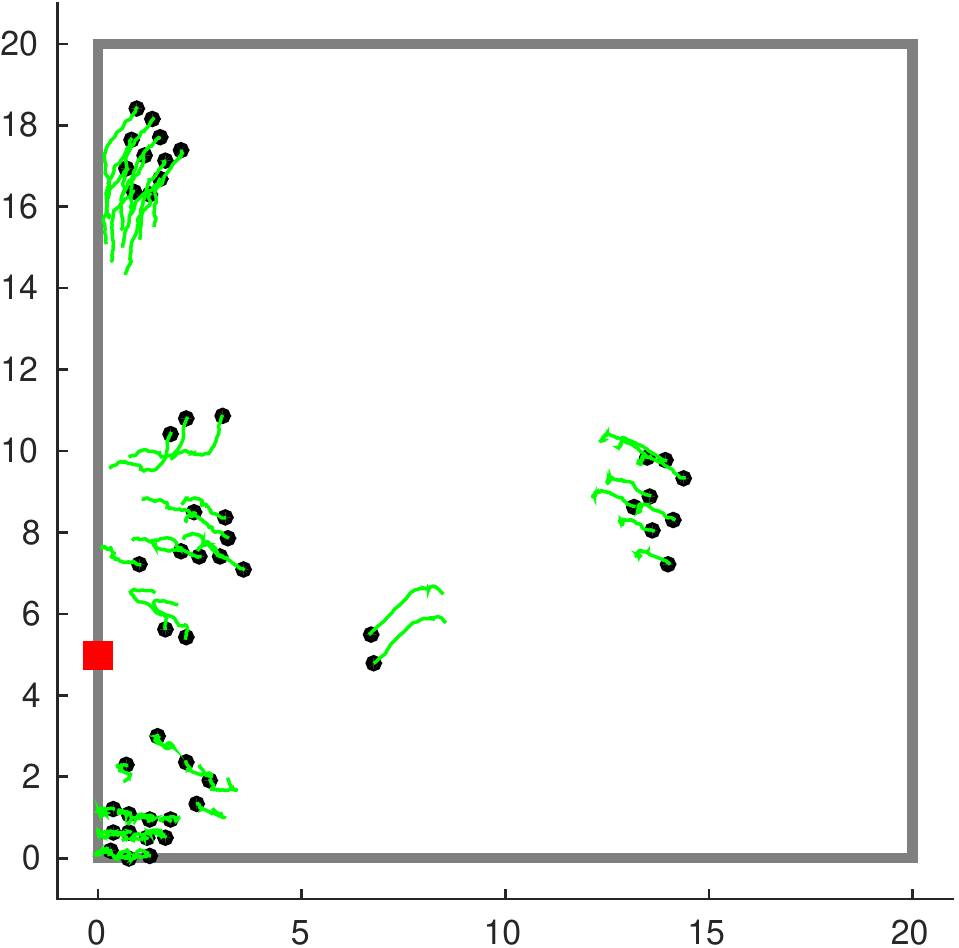}\put(80,80){$P_6$}\end{overpic}
\caption{Screenshots corresponding to the parameters' values $P_1$--$P_6$ in Fig.\ \ref{fig:exploration}. Entrance $\mathcal E$ is depicted as a little red square. No exits and no obstacles are present.}
\label{fig:explorationscreenshots}
\end{figure}

In Fig.\ \ref{fig:explorationscreenshots}, we show instead some screenshots of the simulations corresponding to the parameters' values  $P_1$--$P_6$ of Fig.\ \ref{fig:exploration}.
Choice $P_1$ leads to uncorrelated pedestrians who move mainly in a straight direction (until the boundary is hit), while $P_2$ corresponds to a Brownian-like motion.
Choice $P_3$ leads to strongly cohesive group which is not able to find a consensus about the direction to follow, and then it remains basically fixed, while $P_4$ gives a strongly cohesive group which moves mainly in a straight direction (when the boundary is hit the group finds consensus on a new admissible direction).
Choice $P_5$ leads to similar results of $P_4$ but some times the group splits, especially in presence of obstacles. This is the choice we make in the rest of the paper, since it represents a good compromise between the two tendencies. It is also qualitatively in agreement with the results of the experiment described in \cite[Sect.\ 7]{albi2016SIAP}.
Finally, choice $P_6$ leads to the formation of several subgroups moving around with a rather persisting direction.

\subsection{Statistical analysis of the numerical model}
Due to the presence of the random variable $z$ in the model \eqref{eq:model}, the simulator shows a stochastic behavior, and the results provided are not deterministic. This first source of stochasticity must not be confused with the second source of stochasticity of the optimization problem, connected with the variable number of people involved in the simulation. Here we are dealing with this first aspect, and we are going to identify the minimum number of simulations sufficient for producing a statistically significant database for the determination of the time of egress.
As a consequence, the dynamic of the crowd is different each time the simulation is repeated. However, the configuration of the room (location of entrance and exit, shape of the room, presence of obstacles, etc.) is undoubtedly influencing the time to target, but due to the aforementioned stochastic elements the global effects can be evaluated only in a statistical sense.
Therefore, we need to produce a sufficient number of simulations in order to create
a significant statistical base and the number of simulations needs to be selected and
fixed in advance. This choice represents a trade-off between the stability of the running average value of
the time to target (that is, the number of simulations is sufficiently large so that the
average value is stable and it does not change significantly if further simulations are
included), and the feasibility of the solution of the optimization problem (decreasing
with the selected number of simulations for a prescribed configuration of the room). 
In fact, the objective function is computed a large number of times before the optimal configuration
is detected by the optimization algorithm, so that the number of simulations required for a single
evaluation of the time to target cannot be too large, otherwise the full computational time
makes the solution infeasible.

\begin{figure}[h]
\begin{center}
\includegraphics[width=0.95\textwidth]{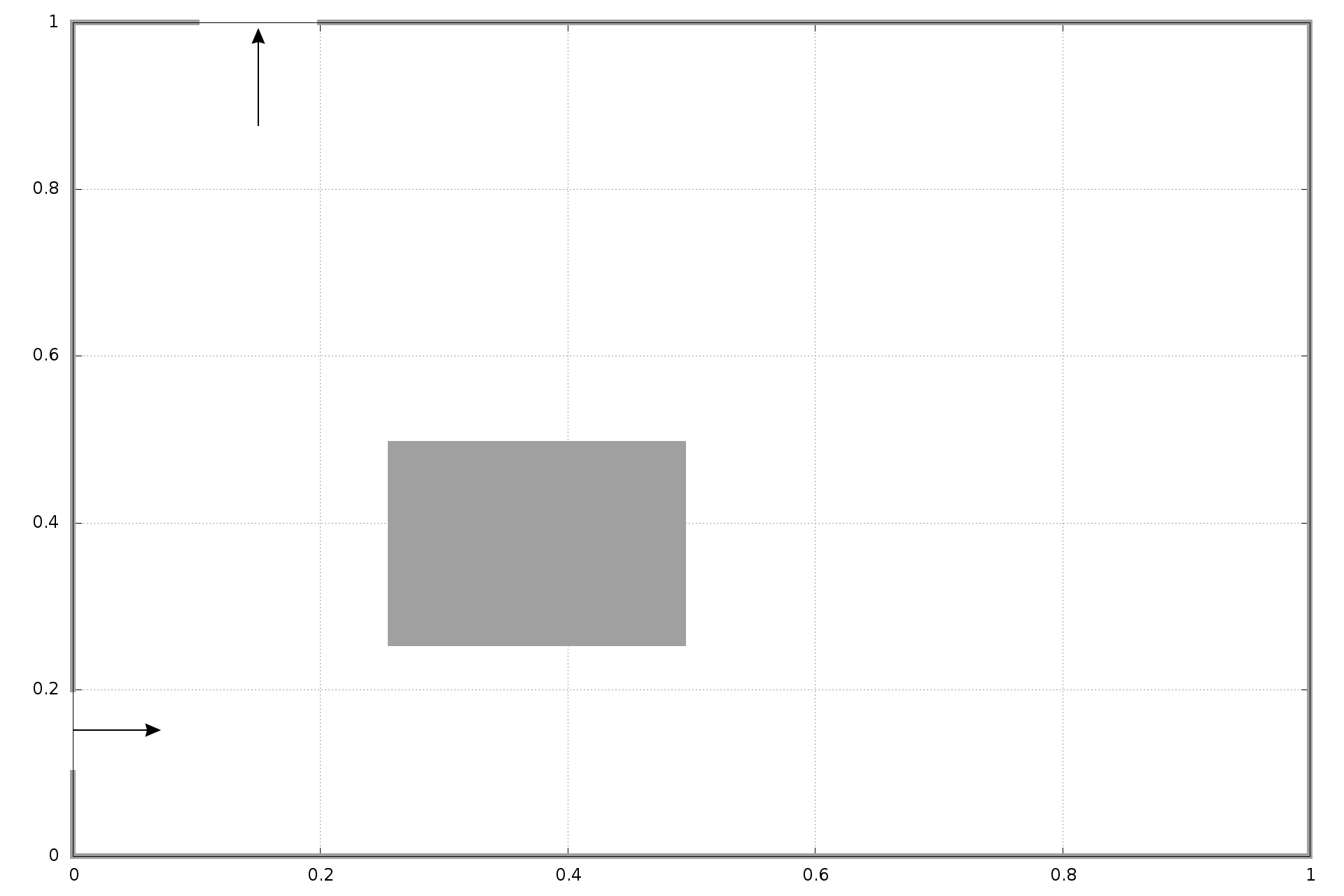}
\caption{Shape of the obstacle considered for the statistical study of the
         simulator. 50 pedestrians are entering a square room with a single exit.
         }\label{Obs}
\end{center}
\end{figure}

To this aim, some preliminary computations have been produced in order to assess a convenient
number of simulations. A simple square room with a single square obstacle has been considered
for this test. The room have a single entrance and a single exit, and 50 pedestrians are observed while
exploring and then leaving the room. A large number of simulations has been performed for this
specific situation, and the results are analyzed statistically. The geometry of the test case
is reported in figure \ref{Obs}: the entrance is located lower left, while the exit is located
upper left. The inclusion of an obstacle is required in order to activate the effects connected
with its presence. In order to be sure that the obstacle is encountered
by the pedestrians, it has been placed in the middle of the room. This is not expected to be the best
configuration, but it is indeed able to put in evidence the role of the interactions between the pedestrians and the obstacle. 

\begin{figure}[h]
\begin{center}
\includegraphics[width=0.95\textwidth]{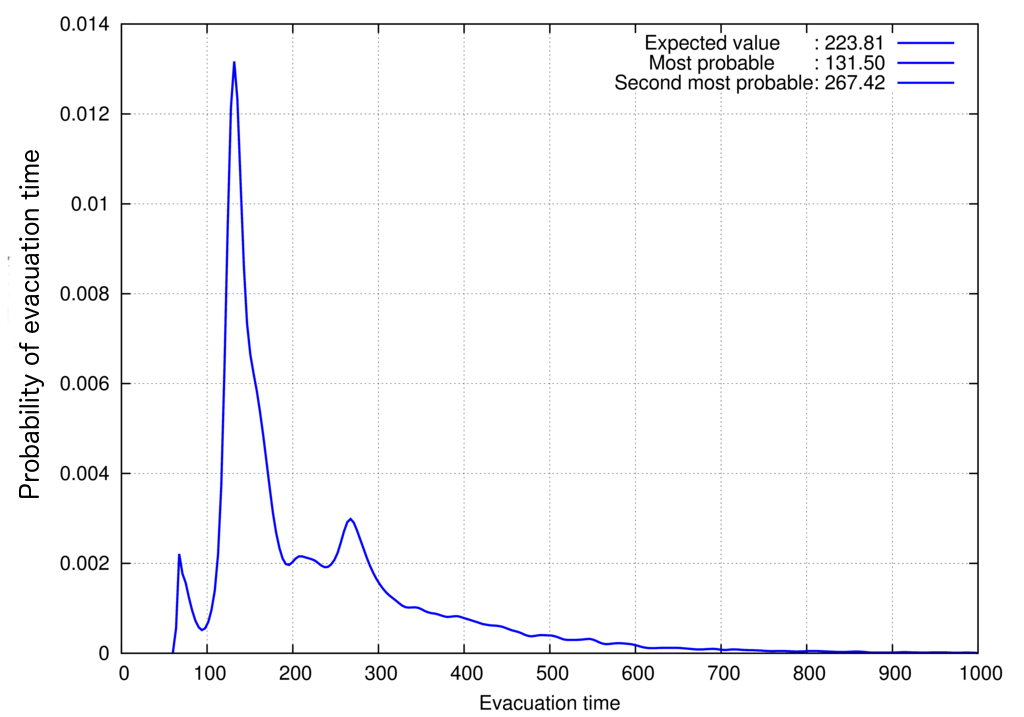}
\caption{Probability Density Function for the time to target.
         50 pedestrians are entering a square room with a single exit
         and a single square obstacle as in figure \ref{Obs}.
         }\label{PDF}
\end{center}
\end{figure}

4,000,000 simulations have been performed for the configuration previously described.
From this database, we derive the Probability Density Function (PDF) of the time to target, reported in figure \ref{PDF}. We can observe how the shape of the PDF is far from a Gaussian curve,
and two different peaks are evident. The first peak, reporting also the highest
value for the probability of occurrence, is indicating a time of 131.50, while the
second peak is found at 267.42. The probability is zero for values lower than the
minimum egress time ever observed (63.67), so that the PDF has a sudden start
from that value on. Expected Value is 223.81, that is far from the most probable one.

\begin{figure}[h]
\begin{center}
\includegraphics[width=0.95\textwidth]{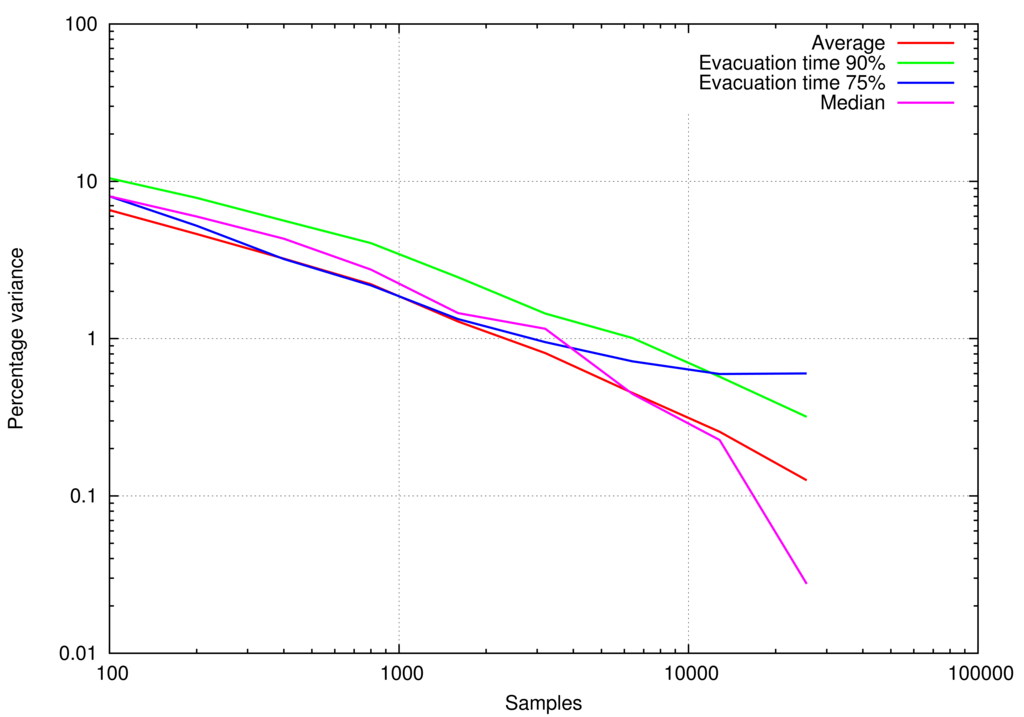}
\caption{Percentage variance associated with time for which, in the 75\% and 90\% of
         the simulations, the crowd has exited the room, reported as a function
         of the number of simulations.
         }\label{SS}
\end{center}
\end{figure}

Now we need to identify a statistical quantity to be adopted as a base element for the objective function of our optimization problem. This quantity will not be the objective function itself, but the objective function will be obtained as a combined analysis of the value of the selected indicator and the PDF. For this reason, it is very important that this indicator is representative of the statistical process. Analyzing the database, we can argue that the average value does not have the required characteristics.
Different tests have been produced for a variety of possible candidates, as partly reported in
figure \ref{SS}, using classical statistical indicators, as also suggested in \cite{ronchi2014},
plus some other quantities, like the time at which, in a prescribed percentage fraction $A$ of
the simulations, all the pedestrians leaved the room (TA). For this last quantity, in figure \ref{SS} the case of 90\% and 75\% are
reported, together with other classical indicators, and 90\% appears the most suitable value. To better
understand how the quantity is computed, we must observe the Cumulative Density
Function (CDF) of the time to target, that is varying from 0 to 1 when the time to target is ranging
from the minimum to the maximum observed value. Our objective function is represented by the time
for which the CDF reaches 0.9, if a probability of 90\% is required. We will refer from now on to this quantity as T90.

\begin{figure}[h]
\begin{center}
\includegraphics[width=0.95\textwidth]{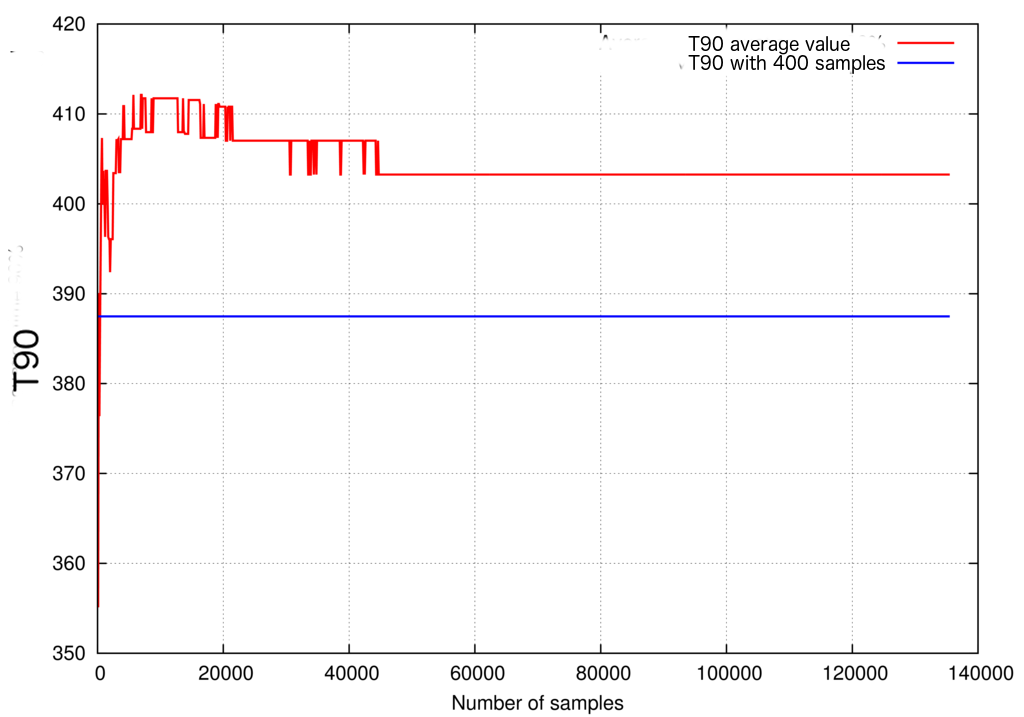}
\caption{Time for which, in the 90\% of the simulations, a group of 50 pedestrians has
         exited the room depicted in \ref{Obs}, reported as a function of the number
         of simulations. The value for the case of 400 simulations is also reported.
         }\label{T90}
\end{center}
\end{figure}

While the average value
can be observed together with its variance, in order to understand the variability of
this quantity with the dimension of the database, T90 represents a simple scalar quantity,
since some statistics are already self-contained. In order to have an idea about the
variability of T90 with the number of simulations, we have computed T90 for an increasing
number of samples, observing its variation as a function the number of simulations adopted.
Results are reported in figure \ref{T90}. Here we can observe that a stable value for T90
is obtained when the number of simulations is about 50,000. This number is absolutely
incompatible with the available computational resources, and it will not be applied in
the practical example. If we assume that the number of simulation adopted is 400, the
inaccuracy is lower than 4\% for this specific configuration.

In order to give a measure of the uncertainties associated with the evaluation of
different quantities, the 4,000,000 samples have been grouped in subsets,
and the average value and its variance have been computed. This way, we can evaluate
the average value also for T90, since we have produced one value of T90 for each independent subset.
Here the running average, the median, T90 and T75 (the same as T90, but with the soil at 75\%)
are considered. Results are reported in figure \ref{SS}. We can observe how there are not
significant differences between the various quantities, and only T75 is not monotonically
decreasing. For this reason, we consider T90 as a significant indicator. This final selection 
substantially represents a conservative approach to the problem: in fact, T90 implicitly provides a guarantee of the success of the egress. 
In order not to increase too much the full computational cost, T90 will be computed
basing on a set of 400 simulations (as previously indicated). This has been estimated to be the maximum possible amount
according to the available computational platform. With this configuration, the evaluation
of a single value of T90 takes from 2 to 10 minutes (wall clock time), depending on the
number of pedestrians. The inaccuracy previously computed is lower than 4\%, and we have to take into consideration this detail when commenting the final results.

\subsection{Choice of the objective function}
\begin{figure}[h]
\begin{center}
\includegraphics[width=0.95\textwidth]{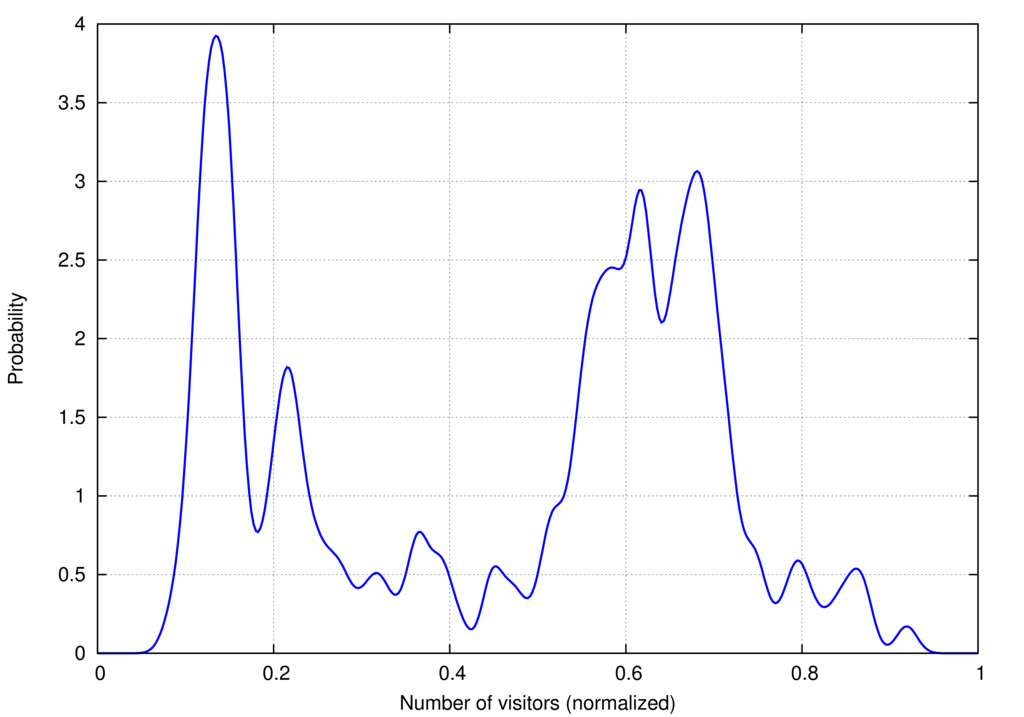}
\caption{Probability Density Function of the number of visitors in the selected archaeological site. 
         On the horizontal axis, the number of pedestrian is rescaled in between the minimum (0)
         and the maximum (1) observed value. On the vertical axis, the flow rate of the visitors
         is reported.
         }\label{Pompei}
\end{center}
\end{figure}

In a previous work \cite{cristiani2017AMM}, the optimization of the shape of one or more obstacles, to be placed
inside a room in order to decrease the time to target of a crowd has
been tackled and solved. In that example, the number of people entering 
the room was fixed. Unfortunately, in real life, the amount of people simultaneously present in a room is almost always unknown and it is rarely constant, cf.\  Section \ref{sec:introduction}. If we consider a cinema, the audience is
depending on the popularity of the movie, the time of the day, the day of the week, etc. Unfortunately,
it is not easy to obtain information about the audience. A practical example,
taken from the web\footnote{{\tt http://www.pompeiisites.org/Sezione.jsp?idSezione=9}}, represents the
situation of the archaeological site of Pompei, Italy. 
Monthly data about the amount
of visitors are publicly available, for a period of about 16 years, so that we can try a statistical
approach to the data. The corresponding PDF is reported in figure \ref{Pompei}. It
represents the probability that a prescribed amount of people is visiting the site in a generic
month. The actual number of people is not available, so that
we cannot observe the number of people concurrently visiting the site in a specific time of
the day: we can only give an estimate on average, assuming that everything is proportional
to the amount of people visiting the site in a month. Anyway, this represents an interesting dataset.

We can observe from figure \ref{Pompei} how two values are substantially the most probable ones,
and the PDF presents a wavy behavior with large variations. Referring to the layout of the previous
examples, we can consider that all the visitors of the archaeological site are supposed to enter the
site passing through the building containing the ticket office, so that the egress from the ticket office to
the site can be reduced at the same type of scenarios previously considered i.e.\ a room
with an entrance and an exit.

Regarding the number of pedestrians visiting the site, we cannot assume a most probable value, since
there are two different values with almost the same probability, and it is also not realistic
to consider the average number of visitors, since it is in between the two most probable ones. The
more convincing approach is to consider the two probabilities together, composing the previously defined T90 with the
PDF of the number of visitors. This way, we have an objective function able
to take into consideration the probability connected with the stochasticity of the egress process
together with the stochasticity of the number of visitors. The final result is an objective function
expressed by
\begin{equation}\label{eq:1d}
  F(\xi) = \frac{\displaystyle\int_{\pi_L}^{\pi_U} T90(\xi,\pi) \cdot p(\pi) \, d\pi}{\pi_U-\pi_L}
\end{equation}
where T90 has been previously defined and depends on the shape of the obstacle inside the room (defined
through the vector of the design variables $\xi$) and the number of pedestrians (denoted by $\pi$), $p(\pi)$ is the probability to have the prescribed amount of visitors $\pi$
into the site, $\pi_U$ is the observed maximum number of pedestrians considered, $\pi_L$ is the observed
minimum number of pedestrians, provided by the experimental data. 

The complexity of the PDF in figure \ref{Pompei} makes the application of low order Gaussian quadrature for the
integration of the objective function practically infeasible. Fortunately, we observe a substantial regularity
of T90 as a function of $\pi$, and we use this characteristics to assume that a spline interpolation can be
substituted to the punctual evaluation of T90 without loss of accuracy. As a consequence, the integration can
be performed with good accuracy, using high order quadrature formulas or even with a simple trapezoidal rule
with a huge number of subdivisions. With this assumption, we are also implicitly neglecting the fact that $\pi$
is an integer variable, so that the integration is not rigorously possible: anyway, if we
assume a large number of pedestrians, we can also reasonably assume that the objective function is not largely
sensitive to the variation of one unit, and its behavior can be considered as continuous. The
interpolation is produced by using five monospaced values over the full interval. 

\begin{figure}[h]
\begin{center}
\includegraphics[width=0.95\textwidth]{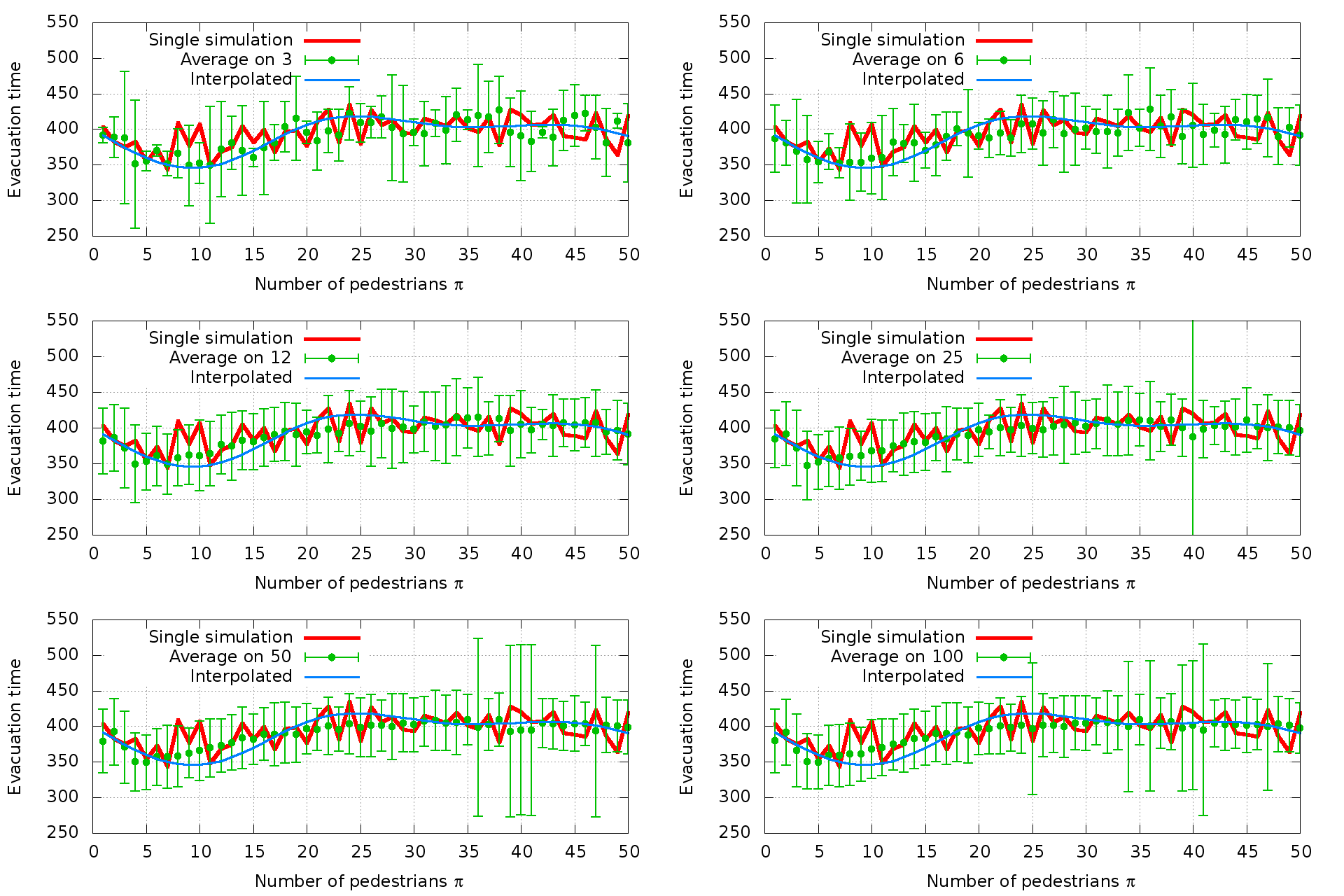}
\caption{Comparison between the real value of the time to target as a function of the number of
         pedestrians and the interpolated value based on the evaluation of the objective function
         in 5 points only. Averaged values (and associated variances) are reported for different 
         statistical bases.
         }\label{Approx}
\end{center}
\end{figure}

In order to give a measure of the accuracy of the interpolation of T90,
the computation of T90 has been repeated from 3 to 100 times for a number of pedestrians in between
2 and 50, and the resulting average values, including the associated variance, are reported in figure \ref{Approx}.
We can see how the interpolated value of the function $T90(\pi)$ lies inside the limits of its
variance. For this reason, the interpolated value can be assumed as representative of the real
one. This is a great result for the simplification of the integration, specifically for the
strong reduction of the overall computational costs.

These preliminary computations allow for some considerations about the computational effort
required by this RDO problem, and some limitations able to make the optimization
problem solvable are obtained from these first experiences. 50 pedestrians has been observed to
be a quite large number if a moderate CPU time is required. For this reason, since the maximum
number of visitors in a month is of about 460,000, a scale factor of 10,000 has been adopted for
the number of pedestrians, so that the integration is performed adopting a number of pedestrians
in the simulation in between 1 and 46. This approximation is absolutely in line with the assumption
that the number of visitors in a month is representative of the amount of people concurrently
visiting the site. Once this last hypothesis is assumed, a further downscale of the size of the
crowd is no further influencing the solution of the problem.

As previously recalled, the interpolation of T90 over $[\pi_L,\pi_U]$ is obtained by distributing five equally-spaced
samples over the definition space. 
After that, the time for the computation of $F(\xi)$ is negligible,
since it is based on the interpolated value of T90 and the computation of the PDF, that is
also based on the interpolation of a limited number of samples.

\section{Shape optimization}\label{sec:shapeoptimization}

Once the objective function has been selected, we have some further elements to be defined in order
to proceed with the determination of the best possible room configuration.

In fact, the optimization problem can be formulated as follows:

\begin{equation}\label{formula}
\min_{\xi\in\R^D} \, F(\xi) \quad \text{subject to}\quad g_i(\xi) \leq 0, \ i=1,\ldots,M,
\end{equation}
where $F(\xi)$ is the objective function of the optimization problem, previously defined in \eqref{eq:1d},
$\xi$ is the vector of the design parameters defining the shape of the obstacle, $D$ is the number of the design parameters, and $g_i(\xi)$ are the $M$ constraint functions for the optimization problem (i.e.\ the conditions that each candidate design must satisfy). 
As previously recalled, the time T90 is a probabilistic function itself, so that a PDF for T90 also exists. A graphical representation has been already
reported in figure \ref{PDF} for a specific configuration.

\subsection{Obstacle's parameterization}

The design parameters $\xi_i$, $i=1,\ldots,D$, are the unknowns of the optimization problem. In this case, we are
looking for the parameters defining the optimal shape of a closed obstacle. As in \cite{cristiani2017AMM},
a B\'ezier curve has been selected for the representation of the obstacle. Reasons are the smoothness
of the curve, the controllability and the reduced number of parameters required. Equation of a generic
B\'ezier curve is
\[
 B(t) = \sum_{i=0}^n \binom{n}{i}P_{i}(1-t)^{n-i}t^i, \qquad t \in [0,1],
\]
where $P_0,\ldots,P_n$ are $n+1$ given points in $\R^2$.

\begin{figure}[h]
\begin{center}
\includegraphics[width=0.95\textwidth]{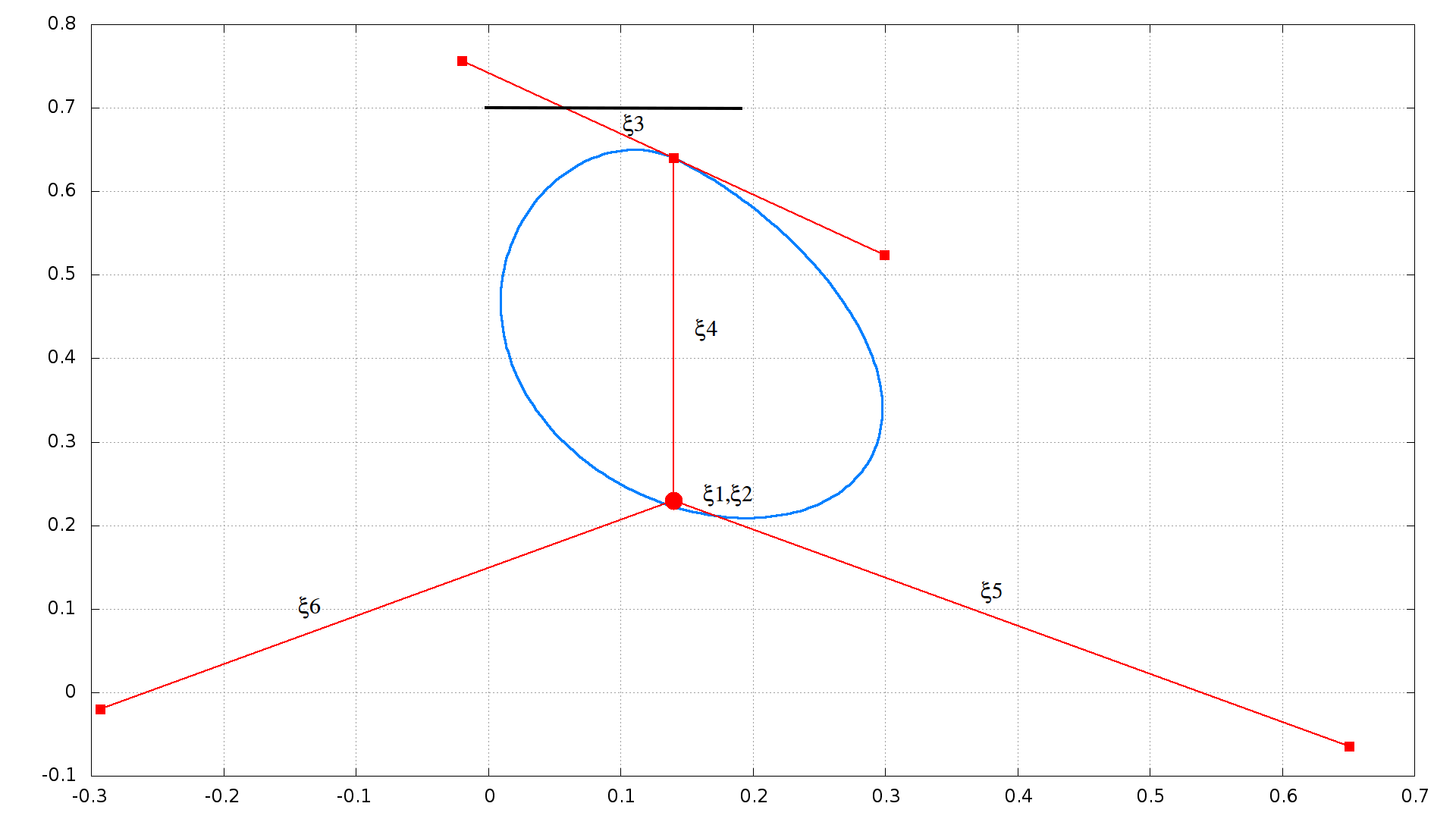}
\caption{Design parameters for the optimization problem. Six free variables 
         define a closed B\'ezier curve.
         }\label{Bez}
\end{center}
\end{figure}

In this case, in order to improve the regularity of the curve, the definition of the parameters is different with
respect to \cite{cristiani2017AMM}. A central point is fixed, and the control points are regularly spaced radially
around the selected center. A last parameter is represented by the tangent at the initial/final point. As a consequence,
if three control points are selected for the B\'ezier curve, six parameters are required, as illustrated in figure \ref{Bez}:
first two parameters $\xi_1$ and $\xi_2$ are fixing the position of the central point, $\xi_3$ is
determining the inclination angle of the tangent at the initial/final point and the remaining parameters
$\xi_4, \xi_5$ and $\xi_6$ are the radial distances of the control points from the central one, being
the angles equally spaced (and consequently already fixed without the necessity of a further parameter).

\subsection{Constraints}

The shape of the obstacle is substantially free, in the limits generated by the selected parameterization. 
However, some limitations are required in order not to produce unrealistic solutions. For this reason, two main
constraints are enforced:

\begin{enumerate}
\item The obstacle cannot occupy the space in proximity of the entrance/exit of the room. The minimum distance is about one meter.
\item The total surface of the obstacle cannot be too small or too large with respect to the room's dimension.
      If we express the surface of the obstacle in percentage points with respect to the full area of the room,
      it must stay in between 20\% and 50\%.
\end{enumerate}

\subsection{Single entrance}

The situation depicted in figure \ref{Obs} has been adopted as the first test case. The rectangular obstacle has
been removed, and a unknown obstacle, parameterized as in figure \ref{Bez}, is considered for the reduction of the
egress time. A single entrance and a single exit are considered in a square room.

The {\tt iPSO} algorithm described in \cite{peri2015EngComp} has been adopted for the solution of the optimization
problem. 
The algorithm is exploring the Design Variable Space (DVS) by spreading a number of agents into this space.
In this case, the coordinates of the DVS are represented by the control points of the B\'ezier curves describing the
obstacle. The set of agents (not to be confused with the pedestrian traced by the simulator) is treated as a swarm,
and each agent is identified by a vector of design parameters.
Basing on the value of the objective function associated with the resulting shape of the obstacle (obtained by applying
the corresponding values of the design parameters), the agents interact and the new speed of each agent is determined
by the simple equation

\begin{eqnarray}\label{eqn:PSO}
v^{k+1}_{i,j}&=&\chi \left [ \omega v^{k}_{i,j} +
 c_{1} \left (p^{k}_{i,j}-x^{k}_{i,j} \right ) +
 c_{2} \left (b^{k}_{  j}-x^{k}_{i,j} \right ) \right ]
\label{eq:velpart} \\
x^{k+1}_{i,j}&=&x^{k}_{i,j}+v^{k+1}_{i,j}
\label{eq:xpart}
\end{eqnarray}

\noindent
where $j=1,2,...,NDV$ indicates the coordinates of the design variable space; $i=1,2,...,n_{sw}$ identifies the agent,
being $n_{sw}$ the total number of agents (typically $4 \times NDV$); $\omega$ is called {\em inertia weight}; $c_{1}$, $c_{2}$ are two positive constants,
called {\em cognitive} and {\em social} parameter respectively; $\chi$ is a {\em constriction factor}, which is used to limit the
velocity; $k=1,2..,K_{max}$ indicates the time step; $p^{k}_{i}$ represents the vector of coordinates of the best position ever
met by the $i^{th}$ agent at time $k$; $b^{k}$ is the vector of coordinates of the best position ever met by the full swarm at
time $k$. The original formulation contains also random coefficients, here removed so as to have a deterministic optimization
algorithm; furthermore, the constriction factor was not present in the original formulation, since it was introduced in \cite{Clerc}.

The time step is applied, and each agent moves to a new position in the DVS. The corresponding value of the objective
function is computed for all the agents, and a new configuration of the swarm is produced. Iteration is repeated until
convergence.
Here the classical PSO algorithm is modified in order to improve convergence and avoid the investigation of unfeasible 
solutions. Convergence is improved by applying a local second order minimizer when the algorithm fails in detecting an improvement
of the current best solution after the movement of the swarm. The computation of the Hessian, needed by the local minimizer,
is strongly simplified by adopting a global approximation model for the objective function (surrogate model \cite{peri2009STR}),
so that no further computations of the objective function are needed. The surrogate model is trained by using the previously
computed values of the objective function. Elimination of unfeasible solutions is obtained by repeating the PSO basic iteration
for those elements falling outside the constrained region of the DVS at the end of the iteration: the iteration is repeated
(for those elements only) until a feasible solution is detected.

\begin{figure}[h]
\begin{center}
\includegraphics[width=0.95\textwidth]{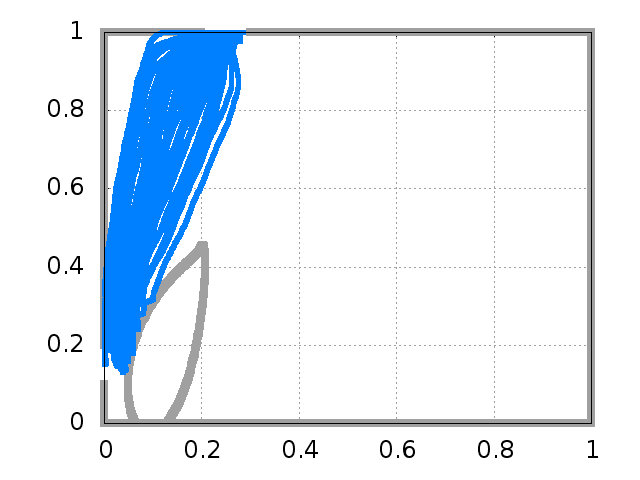}
\caption{Shape of the optimal obstacle obtained after the application of the optimization framework. Trajectories for one simulation are also reported.
}\label{ObsOPT}
\end{center}
\end{figure}

A maximum number of $1000 \times D$ objective function evaluations has been fixed as the
termination criteria for the algorithm. An optimal configuration has been obtained, and its shape
is reported in figure \ref{ObsOPT}. The path followed by the pedestrians is bended by the obstacle
toward the exit, reducing drastically the time to target: an improvement of about 6.4\% is obtained.
Since the improvement is higher than the precision limit of the T90 evaluation, we can conclude that
the optimal solution in undoubtedly better than the original one.

\begin{figure}[h]
\begin{center}
\includegraphics[width=0.95\textwidth]{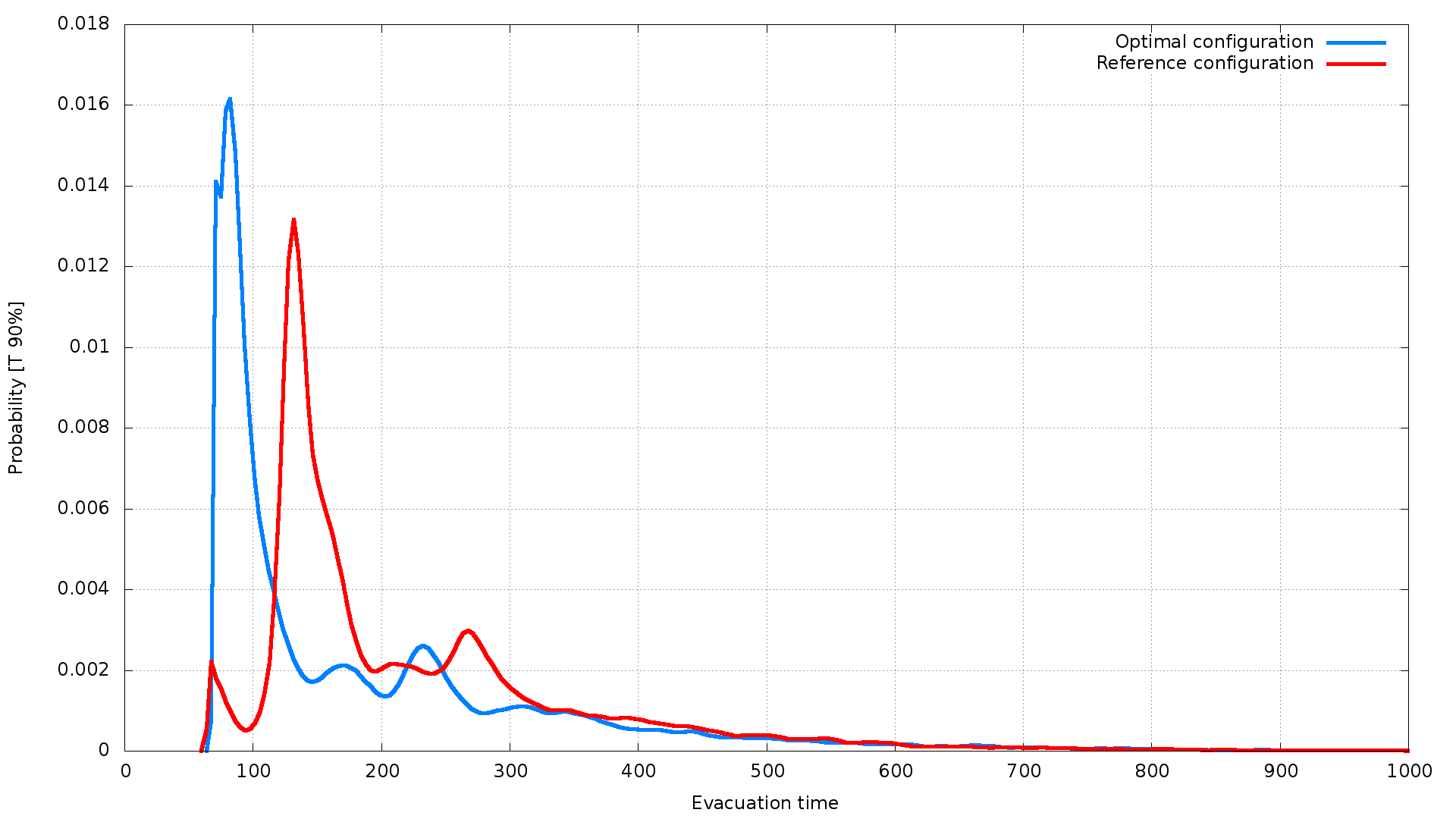}
\caption{Comparison of the Probability Density Functions for the time to target of the room with the optimally
         shaped obstacle and for the reference configuration reported in figure \ref{Obs}.
         }\label{PDFopt2}
\end{center}
\end{figure}

In figure \ref{PDFopt2} the comparison of the PDFs for the reference and the optimal configuration are
reported and compared. The peak of highest probability is shifted backward, with an increase in the
corresponding probability. The minimum value, indicated by the start of the PDF, is almost the same,
while also the time associated with the second peak is reduced.

\begin{figure}[h]
\begin{center}
\includegraphics[width=0.95\textwidth]{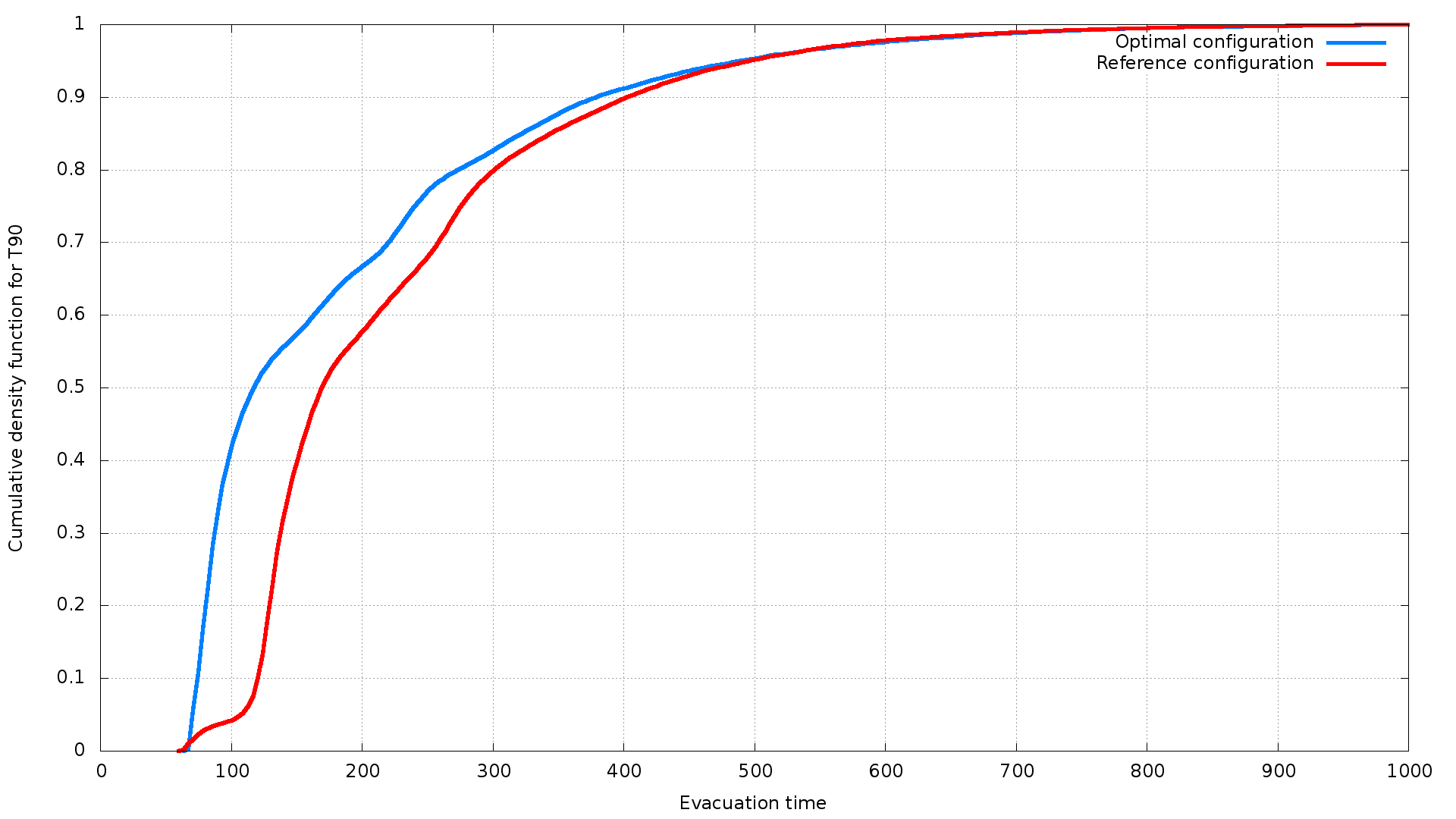}
\caption{Cumulative Density Function for the optimal configuration and for the
         reference configuration depicted in \ref{Obs}.
         }\label{copt}
\end{center}
\end{figure}

The same behavior is obviously observed through the CDF, reported in
figure \ref{copt}. Here is clear how the solutions are much denser in the first part of the curve,
while the very end of the curve presents obviously a very similar behavior.

\begin{figure}[h]
\begin{center}
\includegraphics[width=0.95\textwidth]{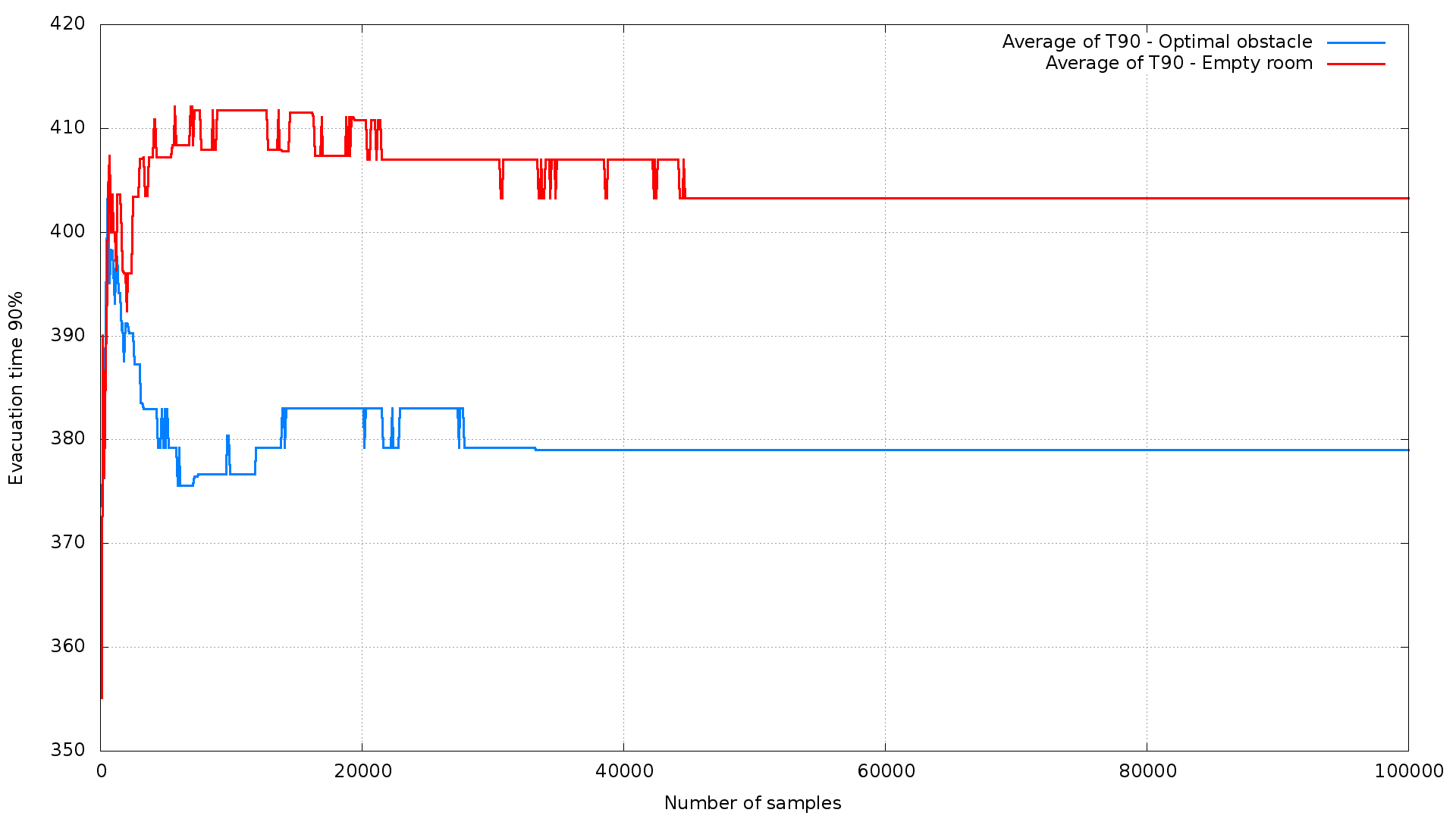}
\caption{Time for which, in the 90\% of the simulations, a group of 50 pedestrians has
         exited the room depicted in \ref{Obs}, reported as a function of the number
         of simulations. Comparison between the case presented in \ref{Obs} and the
         optimal configuration.
         }\label{T90opt}
\end{center}
\end{figure}

A last verification has been produced for the convergent value of T90. The case of 50 pedestrians, already
discussed and reported in figure \ref{T90}, is here repeated for the optimal configuration. Results are shown
in figure \ref{T90opt}. The decrease of T90 is very clear, and also the speed of convergence is increased
for the optimal configuration. This is an indirect proof of the qualities of the selected obstacle, able to
drive the flow of the pedestrians for every variety of initial conditions. With respect to this configuration,
the improvement is of about 6\%. This value is computed using full convergent values of T90, so that we have
here an indirect proof that the order of magnitude of the improvement is confirmed.

\subsection{Multiple entrances}

A second test case has been analyzed, arranging four entrances at the vertexes of a square room
and four exits at the center of each side. The solution has been found applying the same
PDF as in the previous case.

\begin{figure}[h]
\begin{center}
\includegraphics[width=0.95\textwidth]{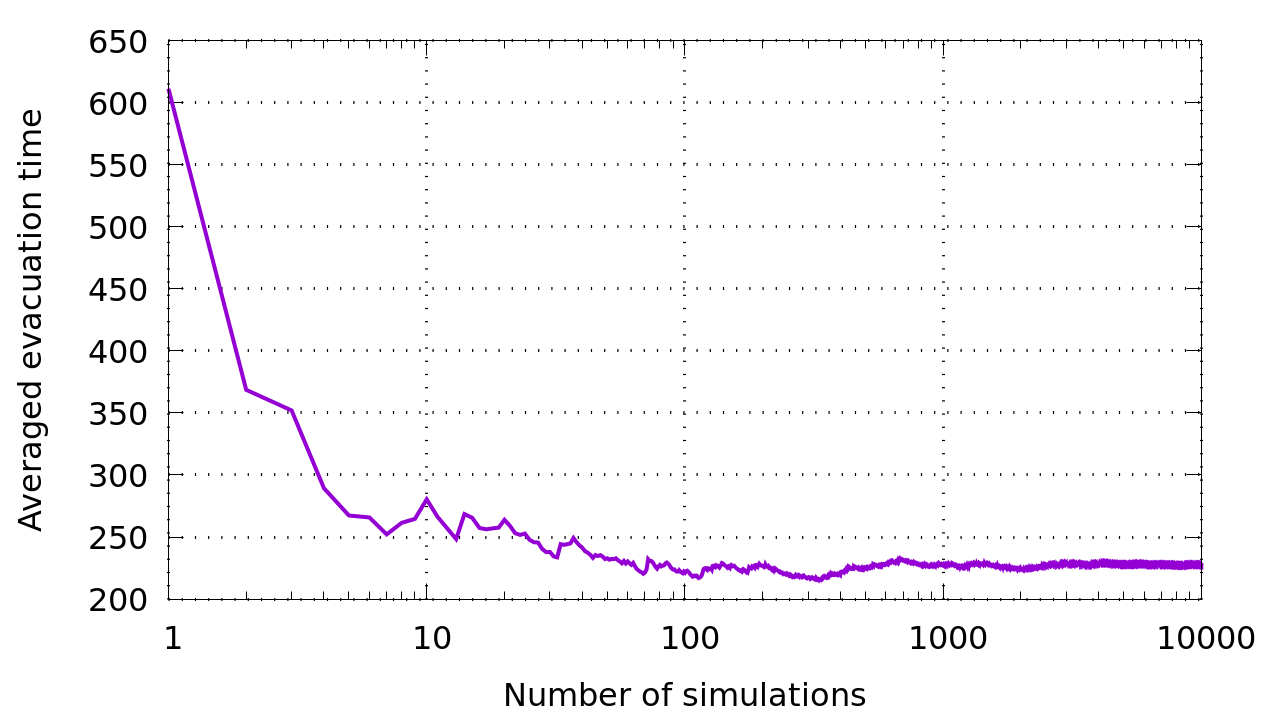}
\caption{Averaged value of the time to target for the configuration with four entrances
         and four exit.
         }\label{Media4}
\end{center}
\end{figure}

Due to the change of the environment, the convergence analysis of the average value of the time to
target has been repeated, and it is reported in figure \ref{Media4}: here the suggested value for the number of
simulation is in between 1,000 and 10,000, and we selected a value of 3,200, remarkably higher than before.

\begin{figure}[h]
\begin{center}
\includegraphics[width=0.95\textwidth]{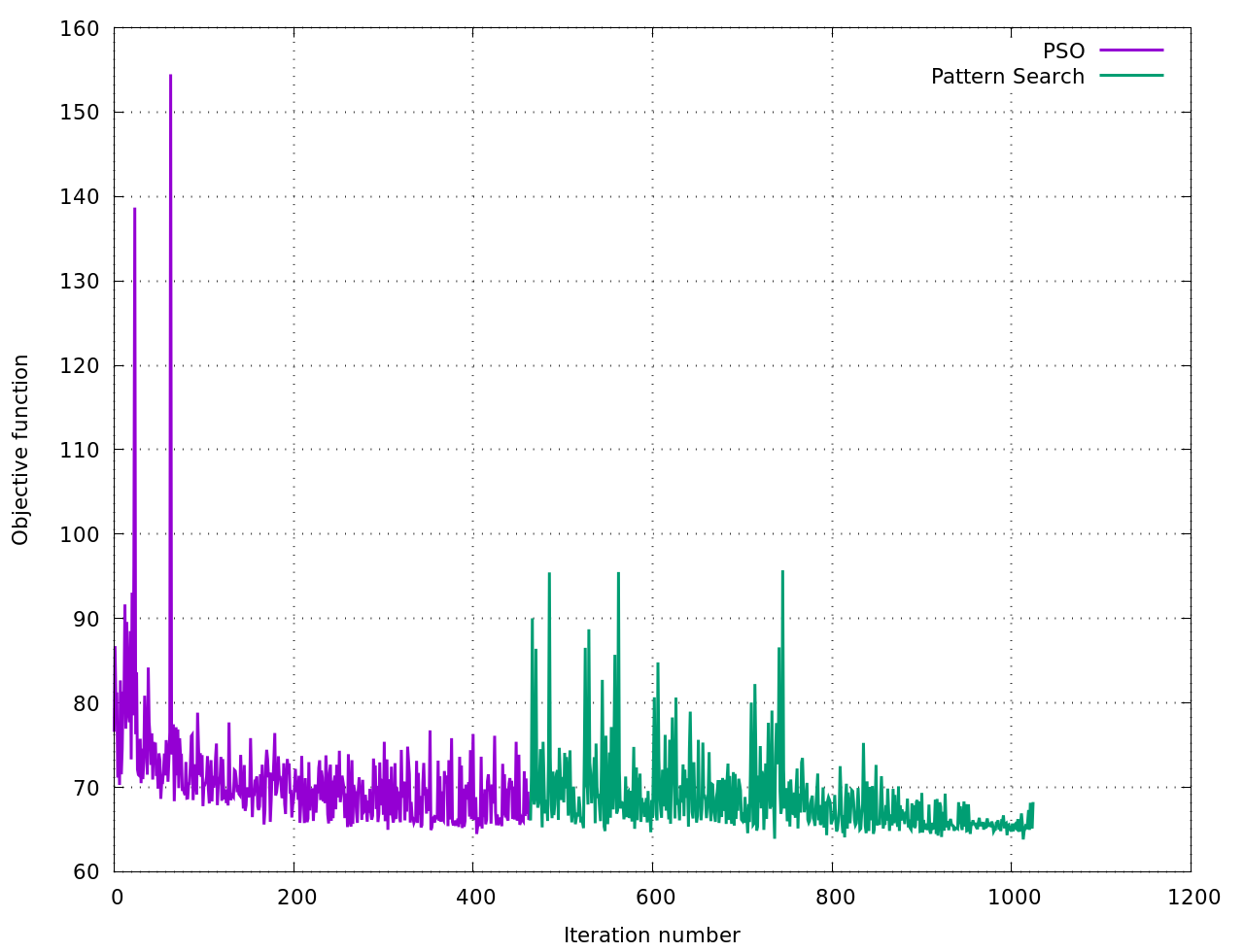}
\caption{Convergence history of the optimization problem. Case of opaque obstacle.
         The first set of solution is driven by the {\tt iPSO} algorithm, while the final
         phase is obtained by using a pattern-search algorithm.
         }\label{Converge}
\end{center}
\end{figure}

Due to the geometrical symmetry of the problem, also the shape of the obstacle can be obtained in four different
ways, since the position and the shape of the obstacle can be regarded as symmetrical with respect to the horizontal
and vertical axes placed at the center of the room. Since this feature has not been considered during the solution of the
optimization problem, the objective function is also cyclical, increasing further the complexity of the problem.
Two different optimization algorithms have been here adopted in sequence: the previously described {\tt iPSO}
algorithm has been truncated at a certain iteration number, and the best solution has been further improved
by a pattern-search algorithm \cite{kolda2004}. The result of this combination is depicted in figure \ref{Converge}:
we can observe that only marginal improvements are obtained by the pattern-search algorithm, confirming the
ability of {\tt iPSO} to identify a good solution.

\begin{figure}[h]
\begin{center}
\includegraphics[width=0.95\textwidth]{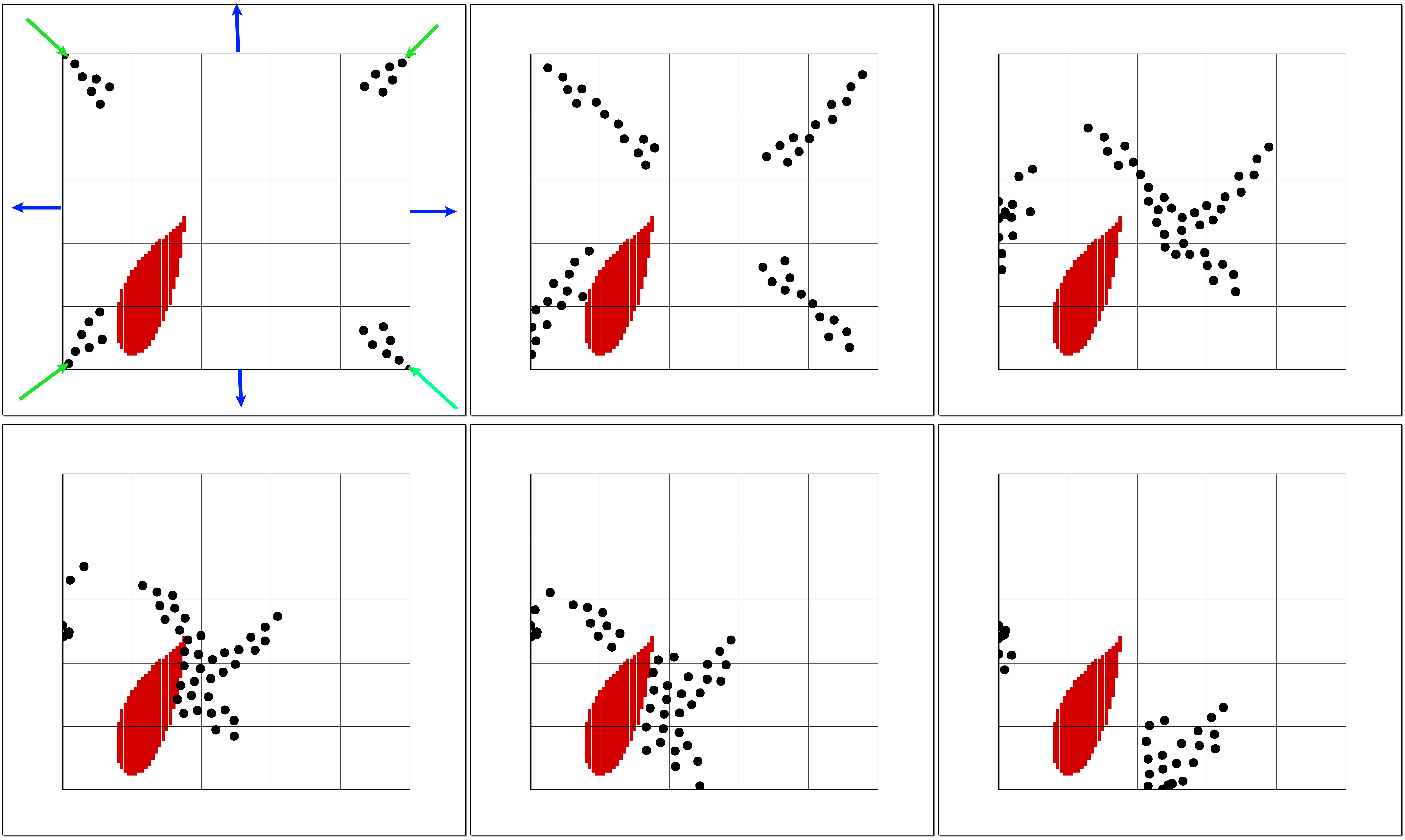}
\caption{Path of 50 pedestrians from a square room with four entrances and four exits. 
Case of opaque obstacle.
         }\label{Opaque6}
\end{center}
\end{figure}

\begin{figure}[h]
\begin{center}
\includegraphics[width=0.95\textwidth]{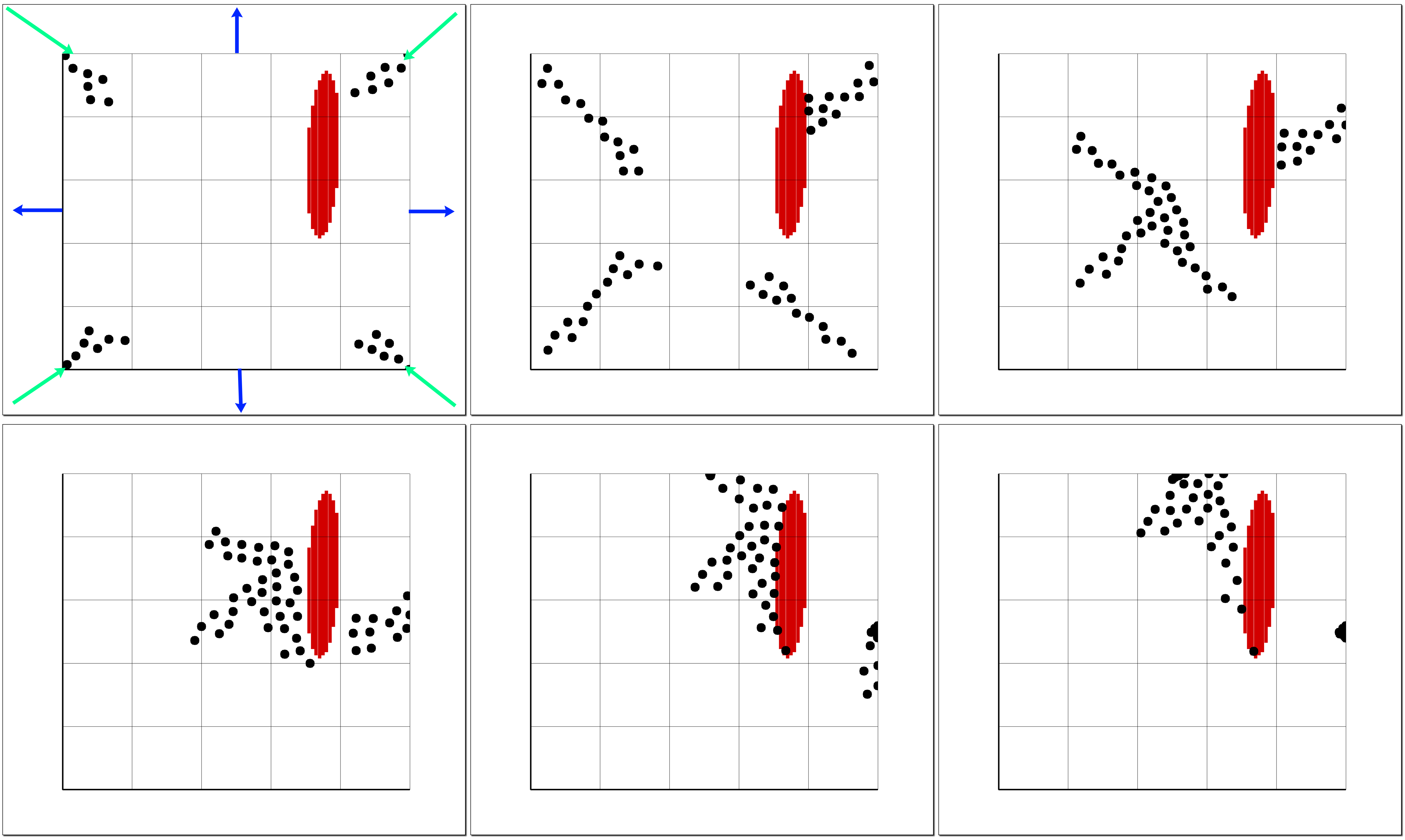}
\caption{Path of 50 pedestrians from a square room with four entrances and four exits. 
Case of transparent obstacle.
         }\label{Transpa6}
\end{center}
\end{figure}

The numerical model has been also modified in order to eliminate, if needed, the transparency of the obstacle. In this case, the pedestrian can or cannot interact with the others on the opposite side of the obstacle. Two
different solutions are expected by the two distinct problems. The dynamic of the crowd is represented in the
pictures \ref{Opaque6} and \ref{Transpa6}.


Location of the obstacles are specular in the two cases: no great differences are observed in this sense.
On the contrary, the shapes of the obstacles are not superimposable but, although different, we can observe
a strong similitude between the two shapes. Firstly, the height-to-width ratio of the obstacles is similar,
and also the position, in front of one entrance, is shared. The obstacle is segregating one part of the crowd,
hiding it to the others and then driving this smaller part to the closets exit. At the same time, the 
other three groups are substantially concurring in the center of the square: this is an effect of the proximity
of two orthogonal walls at the two sides of the entrances. The absence of one fourth of the crowd when the three
groups are meeting at the center of the room \emph{breaks the symmetry} of the configuration. Each group is concurring
with its proper velocity, and the unbalanced component is naturally driving the crowd along the missing direction.
The obstacle, in this case, is repelling the crowd, shifting the flow toward the closest exit, protecting the
segregated part that is using a different exit.

\section{Conclusions}\label{sec:conclusions}
In this paper, we have studied the robust design of an environment which must be left by a unknown number of people in minimum time. 
Additional obstacles are used to \emph{favor} the egress, in the same spirit of other studies about the Braess's paradox in the context of pedestrian flow modeling. 
Robust Design Optimization comes to overcome the main criticism of the previous studies on the design optimization, namely the strong dependence of the optimal displacement of the obstacles on the number and initial positions of the people. 
The adopted technique is able to find a solution whose degree of optimality correctly depends on the probability to observe a certain number of people entering a room.  
The simulations' results give precise suggestions to environment designers: obstacles must block
undesired direction of motion, and, even most important, \emph{block the formation of symmetrical configurations of people}. Indeed, the symmetrical configurations slow down the choice of the direction to follow, since it remains unclear which person has to move first. 
In addition, obstacles can split large groups in order to have better usage of all the exits and avoid congestion in front of the doors, thus benefiting both the first arrivals and, more surprisingly, the late arrivals too.

Some questions arise on the real feasibilty of the optimal shape, that is probably too complex to be implemented in practice.
Here the selection of the {\em Bezier} curves was motivated by the generalty of the searched solution. In a further study, the
obstacle could be obtained as a set of linear walls, in order to prevent unrealistic shapes.

\clearpage

\section*{Acknowledgements}
The authors would like to thank the Italian Minister of Instruction, University and Research (MIUR) to support this research
with funds coming from PRIN Project 2017 (No. 2017KKJP4X entitled "Innovative numerical methods for evolutionary partial
differential equations and applications").

\end{document}